\documentclass[fleqn,usenatbib]{mnras}

\usepackage{newtxtext,newtxmath}

\usepackage[T1]{fontenc}

\DeclareRobustCommand{\VAN}[3]{#2}
\let\VANthebibliography\thebibliography
\def\thebibliography{\DeclareRobustCommand{\VAN}[3]{##3}\VANthebibliography}

\usepackage{graphicx}	
\usepackage{amsmath}	
\usepackage{siunitx}
\usepackage{color}

\title[Atmospheric turbulence profiling with a SH-WFS]{Atmospheric turbulence profiling with multi-aperture scintillation of a Shack-Hartmann sensor}

\author[H.Ogane et al.]{
Hajime Ogane,$^{1}$\thanks{E-mail: h.ogane@astr.tohoku.ac.jp}
Masayuki Akiyama,$^{1}$
Shin Oya$^{2}$
and Yoshito Ono$^{3}$
\\
$^{1}$Astronomical Institute, Tohoku University, 6-3 Aramaki, Aoba-ku Sendai, Miyagi 980-8578, Japan\\
$^{2}$National Astronomical Observatory of Japan, 2-21-1 Osawa, Mitaka, Tokyo 181-8588, Japan\\
$^{3}$Subaru Telescope, National Astronomical Observatory of Japan, 650 North Aohoku Place Hilo, HI 96720, USA
}

\date{Accepted XXX. Received YYY; in original form ZZZ}

\pubyear{2020}

\begin{document}
\label{firstpage}
\pagerange{\pageref{firstpage}--\pageref{lastpage}}
\maketitle

\begin{abstract}
Adaptive optics (AO) systems using tomographic estimation of three-dimensional structure of atmospheric turbulence requires vertical atmospheric turbulence profile, which describes turbulence strength as a function of altitude as a prior information. 
We propose a novel method to reconstruct the profile by applying Multi Aperture Scintillation Sensor (MASS) method to scintillation data obtained by a Shack-Hartmann wavefront sensor (SH-WFS). 
Compared to the traditional MASS, which uses atmospheric scintillation within 4 concentric annular apertures, the new method utilizes scintillation in several hundreds of spatial patterns, which are created by combinations of SH-WFS subapertures. 
Accuracy of the turbulence profile reconstruction is evaluated with Bayesian inference, and it is confirmed that turbulence profile with more than 10 layers can be reconstructed thanks to the large number of constraints. 
We demonstrate the new method with a SH-WFS attached to the 50 cm telescope at Tohoku university and confirm that general characteristics of atmospheric turbulence profile is reproduced.
\end{abstract}

\begin{keywords}
atmospheric turbulence -- scintillation -- Shack-Hartmann wavefront sensor -- adaptive optics
\end{keywords}



\section{Introduction}
Refractive index fluctuation in the Earth's atmosphere distorts the wavefront or equiphase surface of the starlight and causes blur of the stellar image.
Adaptive optics (AO) systems realize diffraction-limited spatial resolution images with ground-based large aperture telescopes. 
In AO systems, with using a natural guide star (NGS) or an artificial laser guide star (LGS) as a reference source, the distortion of wavefront is measured by a wavefront sensor (WFS) and corrected by a deformable mirror (DM) in a timescale of $\sim 1$ millisecond.

In the last decade, in order to improve the performance of an AO system with single LGS, which is affected by the cone effect (\citealp{tallon1990adaptive}) and the angular anisoplanatism (\citealp{stone1994anisoplanatic}), AO systems using multiple LGSs and WFSs have been demonstrated or developed for 8m-class telescopes (e.g. \citealp{marchetti2007sky}, \citealp{arsenault2012eso}, \citealp{lardiere2014multi}, \citealp{rigaut2014gemini}, \citealp{minowa2017ultimate}). 
These systems measure the wavefront distortion in several lines of sight and reconstruct the distortions optimized in the direction of science objects using tomographic estimation, aiming AO correction for lower wavefront error (Laser Tomography Adaptive Optics, LTAO) or wider field of view (Multi Conjugate Adaptive Optics, MCAO, \citealp{beckers1988increasing}, \citealp{rigaut2018multiconjugate}; Ground Layer Adaptive Optics, GLAO, \citealp{rigaut2002ground}, \citealp{tokovinin2004seeing}; Multi Object Adaptive Optics, MOAO, \citealp{hammer2004falcon}, \citealp{vidal2010tomography}). 
The tomographic turbulence estimation is essential technique for next-generation giant segmented mirror telescopes (GSMTs), which have 30m-class primary mirrors.

The tomographic estimation of the three-dimensional turbulence structure requires prior information of the strength of atmospheric turbulence as a function of altitude, which is called atmospheric turbulence profile. 
Tomographic reconstruction matrix is computed from the positions of guide stars and vertical atmospheric turbulence profile. 
Imperfect prior information of the turbulence profile causes tomographic error, which accounts for a large fraction of the total AO error budget \citep{gilles2008wavefront}. 
Because atmospheric turbulence profile varies with time, tomographic reconstruction matrix should be updated in a timescale of tens of minutes, which corresponds to the typical time scale of the profile time evolution (\citealp{gendron2014robustness}, \citealp{farley2020limitations}). 

Altitude resolution is an important parameter for the atmospheric turbulence profiling to reflect the precise turbulence distribution in the tomographic reconstruction matrix. 
\citet{fusco2010impact} and \citet{costille2012impact} studied the effect of the number of layers in turbulence profile on the tomographic error in ELT-scale tomographic AO systems. 
They created less resolved turbulence profile by under-sampling the original highly-resolved (250 layers, $\Delta h \sim 100\ \mathrm{m}$) turbulence profile obtained by a balloon experiment and investigated the impact of the number of layers for reconstruction on the tomographic error. 
The conclusion is that at least 10-20 layers are needed to achieve the tomographic error comparable to that obtained using 250-layer profile. 
The required number of layers depends on LGS asterism diameter and tolerated tomographic error and can be reduced by optimizing the altitude combination of the profile or optimized compression method (\citealp{saxenhuber2017comparison}, \citealp{farley2020limitations}). 

A number of methods to obtain real-time atmospheric turbulence profile based on optical triangulation have been developed. 
Scintillation detection and ranging (SCIDAR; \citealp{rocca1974detection}) uses spatial correlation of scintillation map on the pupil plane from two bright stars with an angular separation of several arcseconds. 
Because variance of intensity fluctuation is proportional to apparent altitude with the power of 5/6, SCIDAR does not have sensitivity to turbulence at low altitudes. 
Generalized-SCIDAR (G-SCIDAR, \citealp{avila1997whole}) has overcome this limitation by detecting scintillation on a plane at some distance away from the pupil plane. 
Slope detection and ranging (SLODAR; \citealp{wilson2002slodar}) has the same triangulation principle with SCIDAR but it uses spatial correlation of phase map on the pupil plane, and the correlation have sensitivity to the ground layer. 
Because these methods are based on optical triangulation between two stars, these methods can be limited by the availability of double stars.
In addition, these methods do not have any sensitivity to turbulence at high altitudes since spatial correlation length created by high turbulence layer is larger than the size of the pupil.

Multi aperture scintillation sensor and differential image motion monitor (MASS-DIMM; \citealp{kornilov2007combined}) is one of the most common profilers, which uses a single star and has lower altitude resolution compared to SLODAR. 
This is a combined method of MASS (\citealp{tokovinin2003restoration}, \citealp{kornilov2003mass}) and DIMM (\citealp{sarazin1990eso}). 
MASS reconstructs 6-layer profile of free-atmosphere based on the scintillation from a single bright star detected by some spatial patterns on the pupil plane.
On the other hand, DIMM measures total seeing or Fried parameter based on the variance of differential image motion through two small apertures. 
Ground-layer seeing is estimated with the difference between the MASS seeing (free-atmosphere seeing) and the DIMM seeing (total seeing). 
Then, 7-layer (ground-layer plus 6 MASS layers) turbulence profile can be obtained. 

Recently, in order to improve the altitude resolution of MASS, fine spatial sampling of scintillation using fast and low-noise detector has been demonstrated by Full Aperture Scintillation Sensor (FASS; \citealp{guesalaga2016fass}, \citealp{guesalaga2020fass}) method.
FASS measures angular power spectrum of scintillation on the pupil plane and compare it with simulated spectra to reconstruct a turbulence profile with 14 layers from 0.3 km to 25 km above the ground.

In this paper, we propose another new turbulence profiling method, which carries out scintillation measurement similar to MASS using spot brightness fluctuation data of Shack-Hartmann wavefront sensor (SH-WFS). 
Hereafter we call this new method SH-MASS.
Compared to the traditional MASS instrument, which measures scintillation with 4 concentric annuli, SH-WFS can measure the scintillation with many spacial patterns created from combinations of subapertures of SH-WFS. 
Thanks to the larger number of measurements, SH-MASS approach utilizes more constraints on the turbulence profile than MASS. 
These constraints make it possible to estimate the atmospheric turbulence profile with high altitude resolution by the observation of scintillation of a single star. 
As an additional merit of using SH-WFS, it is worth noting that SH-WFS can be combined with above-mentioned slope-based profiling methods. 
For example, DIMM can be conducted using differential image motions of two SH-WFS spots and SLODAR can also be executed simultaneously if two SH-WFSs are available. 

This paper is organized as follows. 
In section 2, the principle of MASS is reviewed and application of the method to SH-WFS data is explained. 
In section 3, the response function of SH-MASS is evaluated through simulations with a single layer turbulence. 
In section 4, on-sky experiment conducted with 50cm telescope at Tohoku university is explained and the results of atmospheric profiling are shown. 
In section 5, remaining problems and limitations of SH-MASS are discussed. 
Finally, the whole contents are summarized in section 6.

\section{Principle}
\subsection{Review of the principle of MASS}
\label{sec:principle} 

In this subsection, the principle of MASS is summarized following \citet{kornilov2003mass} and \citet{tokovinin2003restoration}.
Let us consider that light with a wavelength of $\lambda$ passes through a single layer of atmospheric turbulence whose altitude is $h$ and thickness is $\Delta h$.
The phase of the light is distorted by refractive index fluctuation in the turbulence layer according to the turbulence strength which is indicated by structure constant of refractive index fluctuation $C_N^2(h)$. 
Then, assuming the Kolmogorov's atmospheric turbulence model, the spatial power spectrum of the phase fluctuation $\Phi_\phi\mathrm{[m^2]}$ is written as follows,
\begin{align}
    \Phi_\phi(f_x,f_y) = 0.38 \lambda^{-2}f^{-11/3}C_N^2(h)\Delta h,
\end{align}
where $f_x,f_y$ are spatial frequencies and $f=\sqrt{f_x^2+f_y^2}$.
Besides, by assuming the Fresnel propagation, and the weak perturbation (Rytov's) approximation, which is applicable for astronomical observation at not very large zenith angle $z$, the spatial power spectrum of the intensity fluctuation $\Phi_I\mathrm{[m^2]}$ is written as follows,
\begin{align}
    \Phi_I(f_x,f_y) = 4\sin^2\left(\pi\lambda h\ \sec(z) f^2\right)\Phi_\phi(f_x,f_y).
    \label{eq:ps_intensity}
\end{align}
Here, $h\sec(z)$ represents apparent altitude of the turbulence layer, or light propagation distance from the layer.
Generally, multiple turbulence layers at different altitude and thickness affect the intensity fluctuation.
Total contribution from these multiple layers can be written as linear combination of that from each layer,
\begin{align}
    \Phi_I(f_x,f_y) = \sum_{i}^{N_\mathrm{layer}} 1.53f^{-11/3} \left\{ \frac{\sin(\pi\lambda h_i \sec(z) f^2)}{\lambda} \right\}^2 C_N^2(h_i)\Delta h_i,
    \label{eq:ps_intensity2}
\end{align}
where $N_{\mathrm{layer}}$ is the number of layers and $i$ is the index of each turbulence layer.

\begin{figure}
	\includegraphics[width=\columnwidth]{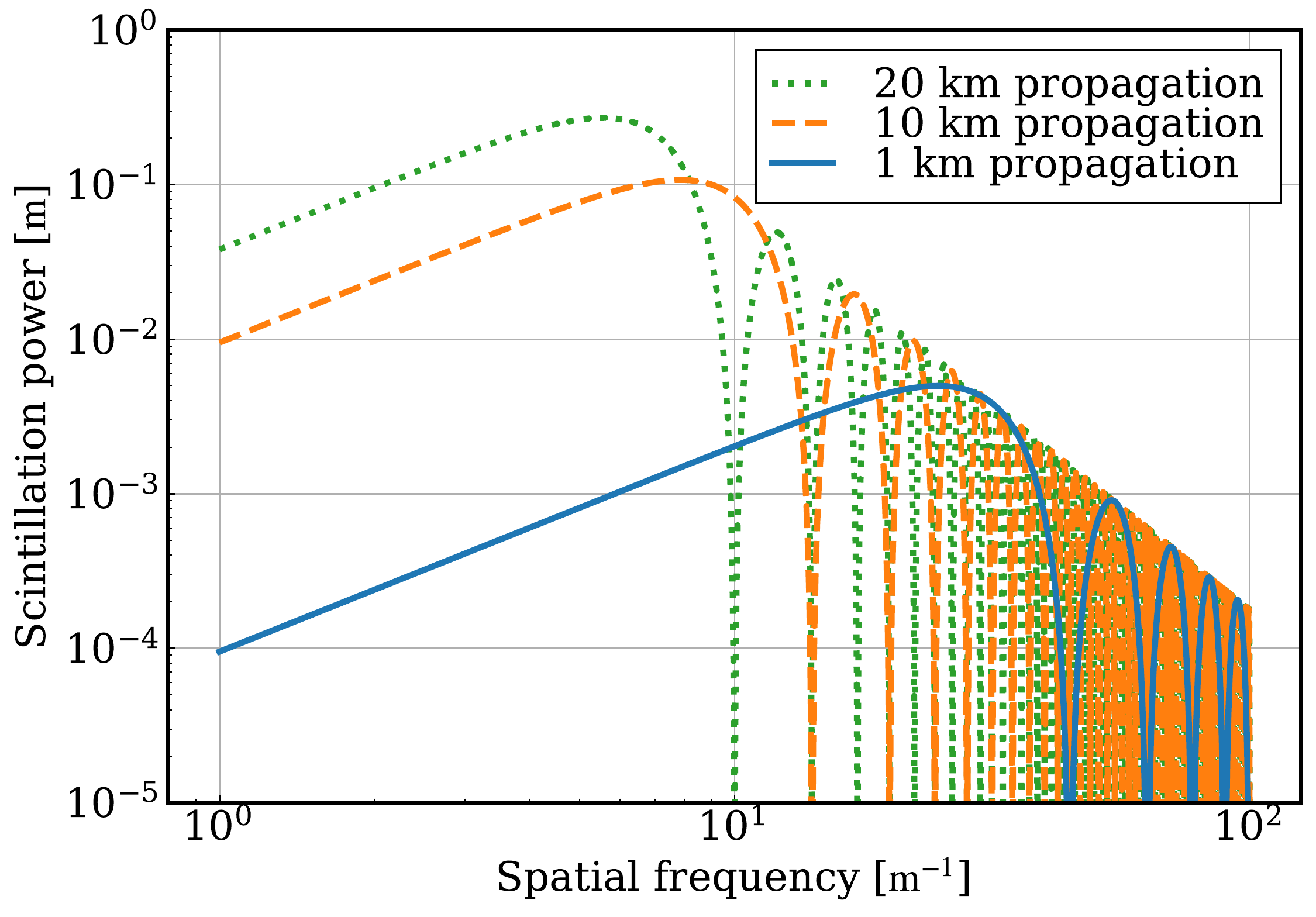}
    \caption{One-dimensional spatial power spectrums of scintillation are shown. Vertical axis means $\Phi_I(f_x,f_y)$ integrated at constant $f$; $2\pi f\Phi_I(f_x,f_y)$. Monochromatic ($\lambda=500\mathrm{nm}$) scintillation power spectrum which is created by a single turbulence layer with a constant turbulence strength ($C_N^2(h)\Delta h = 1.0E-12 \mathrm{[m^{1/3}]}$) at the propagation distance of} 1km, 10km, and 20km is shown as blue solid, orange dashed, and green dotted line, respectively.
    \label{fig:ScintillationPowerSpectrum}
\end{figure}

Fig.\ref{fig:ScintillationPowerSpectrum} shows the power spectrum of intensity fluctuation. 
In this plot, $\Phi_I(f_x,f_y)$ in Eq.\ref{eq:ps_intensity} integrated at constant $f$, i.e. $2\pi f\Phi_I(f_x,f_y)$ is shown as a function of $f$.
Single turbulence layer with a constant turbulence strength and monochromatic ($\lambda=500\ \mathrm{nm}$) observation are assumed.
The spatial frequency at which the power of scintillation has its peak depends on the propagation distance of the turbulence layer.
The layer is higher, the lower spatial frequency accounts for large amount of the power. 
Hence, detecting scintillation at different spatial frequencies makes it possible to discern the contributions from turbulence layers at different altitudes.  
The amplitude of the power of scintillation also depends on the altitude of the turbulence layer, and contributions from lower layers are weaker. 
Scintillation hardly has the information of turbulence at ground layer or dome seeing because $\Phi_I$ in Eq.\ref{eq:ps_intensity} becomes 0 at $h=0$.

MASS is a technique to estimate the turbulence profile by utilizing the dependence of the spatial frequency of the intensity fluctuation on the apparent altitude of the turbulence layer. 
MASS instrument divides the pupil into several concentric annuli (see Fig.\ref{fig:shmass_aperturepattern}) and measures the starlight intensity in the concentric apertures. 
The intensity fluctuation is characterized with the scintillation index (SI), which is the variance and covariance of the normalized intensity observed by the concentric apertures. 
Denoting the observed intensity in the X-th annulus as $I_X$, the intensity variance of the X-th annulus, referred as normal scintillation index, is defined as follows,
\begin{align}
    s_X     &= \mathrm{Var}\left[ \frac{I_X}{\langle I_X \rangle} \right],
    \label{eq:nsi}
\end{align}
where $\langle \rangle$ represents time-average, and $\mathrm{Var}$ means variance. 
Likewise, the intensity covariance of the X-th and Y-th annuli, referred as a differential scintillation index, is defined as 
\begin{align}
    s_{XY} &= \mathrm{Var}\left[ \frac{I_X}{\langle I_X\rangle}-\frac{I_Y}{\langle I_Y\rangle } \right] \nonumber \\
    &= s_X + s_Y - 2\mathrm{Cov}\left[ \frac{I_X}{\langle I_X\rangle },\frac{I_Y}{\langle I_Y\rangle } \right],
    \label{eq:dsi}
\end{align}
where $\mathrm{Cov}$ means covariance. 

These SIs can be expressed using the power spectrum of intensity fluctuation $\Phi_I(f_x,f_y)$ as follows,
\begin{align}
    s_X &= \iint \Phi_I(f_x,f_y) |\mathcal{F}[A_X(x,y)]|^2 df_xdf_y, 
    \label{eq:nsi_psinteg} \\
    s_{XY} &= \iint \Phi_I(f_x,f_y) |\mathcal{F}[A_X(x,y)-A_Y(x,y)]|^2 df_xdf_y,
    \label{eq:dsi_psinteg}
\end{align}
where $\mathcal{F}$ means Fourier transformation and $A(x,y)$ is the normalized aperture function; a function which returns value of 1 divided by the area of the aperture for $(x,y)$ inside the aperture and value of 0 for others. 
By substituting Eq.\ref{eq:ps_intensity2} into $\Phi_I(f_x,f_y)$ in Eq.\ref{eq:nsi_psinteg} and Eq.\ref{eq:dsi_psinteg}, following formula are obtained.
\begin{align}
    s_{X} &= \sum_{i}^{N_\mathrm{layer}} W_{X,i} J_i,
    \label{eq:nsi=wj}\\
    s_{XY} &= \sum_{i}^{N_\mathrm{layer}} W_{XY,i} J_i,
    \label{eq:dsi=wj}
\end{align}
where
\begin{align}
    W_{X,i} &= \iint 1.53f^{-11/3} \left\{ \frac{\sin(\pi\lambda h_i \sec(z) f^2)}{\lambda} \right\}^2 |\mathcal{F}[A_X(x,y)]|^2 df_xdf_y,
    \label{eq:nwf}\\
    W_{XY,i} &= \iint 1.53f^{-11/3} \left\{ \frac{\sin(\pi\lambda h_i \sec(z) f^2)}{\lambda} \right\}^2 \nonumber \\ 
    &\qquad\qquad\qquad\qquad |\mathcal{F}[A_X(x,y)-A_Y(x,y)]|^2 df_x df_y.
    \label{eq:dwf}\\
    J_i &= C_N^2(h_i)\Delta h_i.
\end{align}
$W_{X,i}$ and $W_{XY,i}$ are called normal weighting functions (WFs) and differential WFs, respectively, which can be calculated from the information of aperture geometry and measurement wavelength. 
By solving Eq.\ref{eq:nsi=wj} and Eq.\ref{eq:dsi=wj}, turbulence strengths $J_i = C_N^2(h_i)\Delta h_i$ of multiple layers are estimated. 
Typical MASS instrument has four concentric annular apertures whose diameters are 2.0, 3.7, 7.0, and 13.0 cm, respectively. 
Then, ten SIs (four normal SIs plus six differential SIs) are derived in order to reconstruct a turbulence profile of 6 turbulence layers (0.5, 1, 2, 4, 8, and 16 km above the aperture). 
Each SI is calculated every minute from the photon counting with $\sim 1\ \mathrm{kHz}$ (reference : \citealp{kornilov2003mass}).

In real measurement, spectral characteristics such as spectral energy distribution of the observed star, filter transmission characteristic and detector sensitivity need to be considered. 
According to \citet{tokovinin2003polychromatic}, the effect from the polychromatic scintillation can be described by replacing the WFs as follows.
\begin{align}
    W_{X,i} &= \iint 1.53f^{-11/3} \left\{ \int \frac{\sin(\pi\lambda h_i \sec(z) f^2)}{\lambda} F(\lambda) d\lambda \right\}^2 \nonumber \\
    &\qquad\qquad\qquad\qquad
    |\mathcal{F}[A_X(x,y)]|^2 df_xdf_y,
    \label{eq:nwf_poly}\\
    W_{XY,i} &= \iint 1.53f^{-11/3} \left\{ \int \frac{\sin(\pi\lambda h_i \sec(z) f^2)}{\lambda} F(\lambda) d\lambda \right\}^2 \nonumber \\ 
    &\qquad\qquad\qquad\qquad |\mathcal{F}[A_X(x,y)-A_Y(x,y)]|^2 df_x df_y,
    \label{eq:dwf_poly}
\end{align}
where $F(\lambda)$ is normalized spectral function which contains all the spectral characteristics mentioned above and is normalized to satisfy $\int_{-\infty}^{\infty} F(\lambda) d\lambda = 1$. 

\subsection{Application of MASS to SH-WFS data}
\label{sec:apply_MASS_to_SHWFS}
Because SH-WFS effectively divides the entrance pupil into grid pattern subapertures, it can be used to measure scintillation in many spatial patterns by multiple combinations of subapertures. 
Therefore, SH-WFS is applicable for conducting MASS. 

\begin{figure}
	\includegraphics[width=\columnwidth]{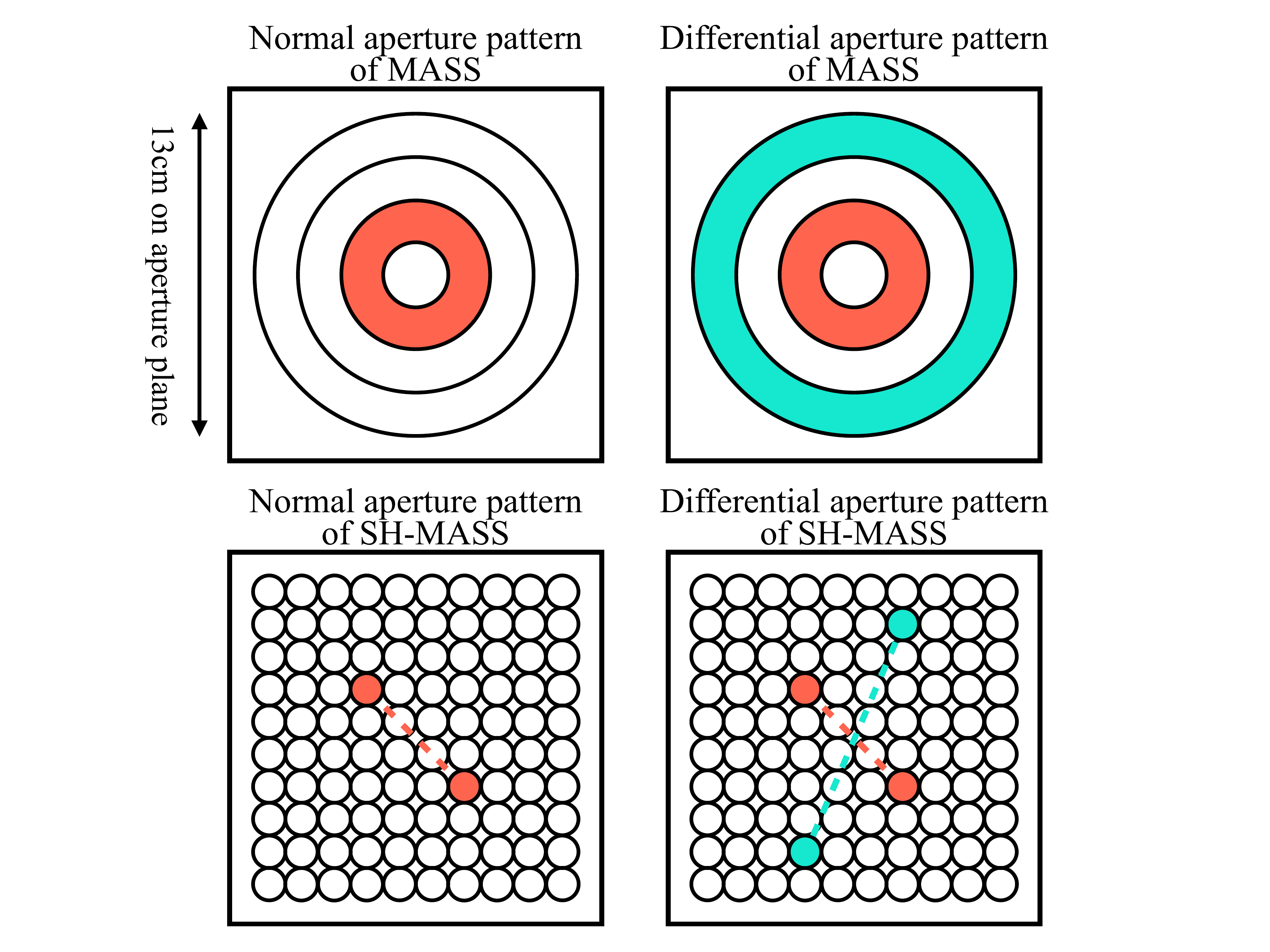}
    \caption{In the top panels, examples of MASS spatial pattern whose measurements are used to compute SIs are shown. For example, a normal SI is calculated as the intensity variance measured by red annular aperture while a differential SI is calculated as the intensity covariance between the measurement by red aperture and that by blue aperture. In bottom panels, examples of spatial pattern for SH-MASS are shown. In our definition, a normal SI is calculated as the intensity variance observed by red subapeture pair while a differential SI is computed as the intensity covariance between the observation by red subaperture pair and blue subaperture pair. In the calculation of differential SIs, only two subaperture pairs which have common mid point are used.}
    \label{fig:shmass_aperturepattern}
\end{figure}

Fig.\ref{fig:shmass_aperturepattern} shows comparison of the definition of spatial patterns in traditional MASS (top two panels) and SH-MASS (bottom two panels). 
In traditional MASS case, concentric annular spatial patterns are used in order to extract scintillation at a specific frequency which corresponds to the diameters of the annulus. 
For example, a normal SI is defined as a intensity variance measured by the red annulus in the left top panel of Fig.\ref{fig:shmass_aperturepattern} while a differential SI is defined as a intensity covariance between the red and blue annulus in the right top panel of Fig.\ref{fig:shmass_aperturepattern}.

On the other hand, we define a subaperture pair, which consists of two subapertures, as one spatial pattern of SH-MASS so that we can effectively extract a certain spatial frequency component of scintillation which is characterized by the distance of the two subapertures. 
Then, total intensity of a subaperture pair is used as a measured value to calculate a SI. 
For example, a normal SI is defined as a fluctuation variance of total intensity measured by the two red subapertures in the left bottom panel of Fig.\ref{fig:shmass_aperturepattern}. 
Based on this definition of normal SI, Eq.\ref{eq:nsi} can be rewritten as,
\begin{align}
    s_X     &= \mathrm{Var}\left[ \frac{I_X}{\langle I_X\rangle } \right] \nonumber \\
    &= \mathrm{Var}\left[ \frac{I_i + I_j}{\langle I_i + I_j\rangle } \right] \nonumber \\
    &= \frac{\mathrm{Var}[I_i]+\mathrm{Var}[I_j]+2\mathrm{Cov}[I_i,I_j]}{(\langle I_i\rangle  + \langle I_j\rangle )^2},
    \label{eq:nsi_spotstat}
\end{align}
where $i,j$ are indices of subapertures which constitutes aperture $X$, and $I_i$ represents spot intensity (or counts) observed in i-th subaperture.
Thanks to a large number of SH-WFS subapertures, there are many subaperture pairs which have common separation distance. 
Then, we calculated normal SIs for all subaperture pairs which have a common spatial distance and regarded average and standard deviation as a normal SI and its measurement error, respectively. 

Likewise, a differential SI is defined as a fluctuation covariance between total intensity measured by the two red subapertures and that measured by the two blue subapertures in the right bottom panel of Fig.\ref{fig:shmass_aperturepattern}. 
Then, Eq.\ref{eq:dsi} can be rewritten as,
\begin{align}
    & s_{XY} = s_X + s_Y - 2\mathrm{Cov}\left[ \frac{I_X}{\langle I_X\rangle },\frac{I_Y}{\langle I_Y\rangle } \right] \nonumber \\
    & = s_X + s_Y - 2\mathrm{Cov}\left[ \frac{I_i + I_j}{\langle I_i + I_j\rangle },\frac{I_k + I_l}{\langle I_k + I_l\rangle } \right] \nonumber \\
    & = s_X + s_Y \nonumber \\
    & \qquad - 2\frac{\mathrm{Cov}[I_i,I_k]+\mathrm{Cov}[I_j,I_k]+\mathrm{Cov}[I_i,I_l]+\mathrm{Cov}[I_j,I_l]}{(\langle I_i\rangle  + \langle I_j\rangle )(\langle I_k\rangle  + \langle I_l\rangle )}
    \label{eq:dsi_spotstat}
\end{align}
where $i,j$ are indices of subapertures which constitutes aperture $X$, while $k,l$ means indices of subapertures for aperture $Y$.
Here, we calculated differential SIs for two subaperture pairs which have common mid point. 
That corresponds to taking concentric two annuli in the traditional MASS. 
By these definitions of spatial patterns in SH-MASS, 51 normal SIs and 234 differential SIs are obtained with $10\times10$ SH-WFS.

\begin{figure}
	\includegraphics[width=\columnwidth]{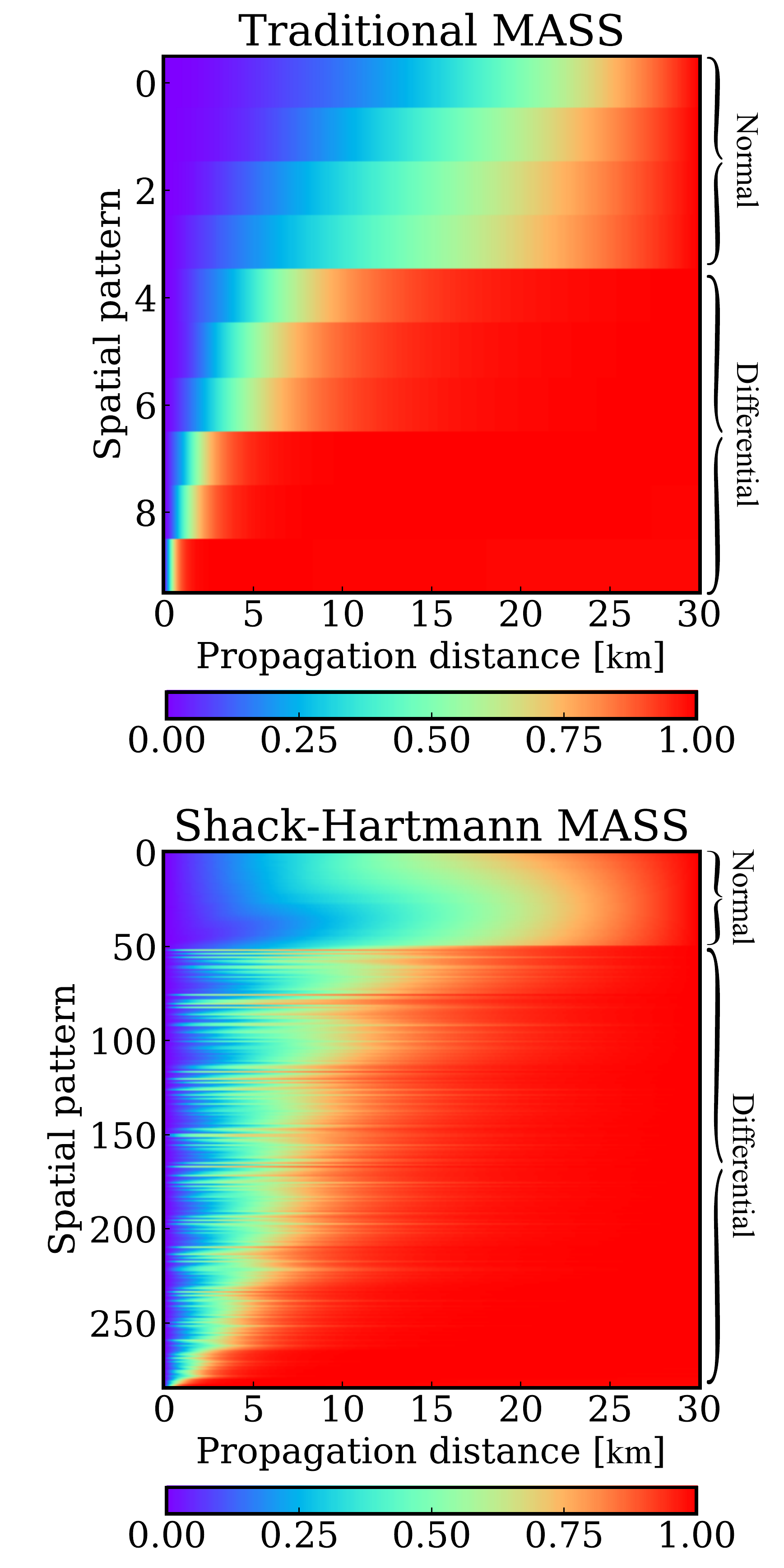}
    \caption{Top: WF matrix of the traditional MASS. Ten spatial patterns are made by four concentric annular apertures whose diameters are 2.0, 3.7, 7.0, and 13.0 cm, respectively. Bottom: WF matrix of the SH-MASS. As the geometry of SH-WFS, $10\times10$ subapertures with each subaperture size scales $1.3\mathrm{cm}$ on the primary mirror was assumed. Both of the WFs are calculated assuming the measured wavelength of 500 $\mathrm{nm}$. Each pixel is coloured with weight value $[\mathrm{m^{-1/3}}]$ normalized in each row so that characteristic propagation distance where weight value reaches 50\%-ile can be seen as light green.}
    \label{fig:WfComp}
\end{figure}

Fig.\ref{fig:WfComp} shows the comparison of WFs of the traditional MASS and SH-MASS which are calculated assuming the aperture geometries shown in Fig.\ref{fig:shmass_aperturepattern} and measurement wavelength of 500 nm.
Each row of the WF matrix represents WF of each spatial pattern, i.e. Eq.\ref{eq:nwf} or Eq.\ref{eq:dwf}. 
Here, values of WFs are normalized in each row so that the weight value transition in the direction of propagation distance can be easily recognized. 
In traditional MASS case, the number of spatial patterns is only ten (four normal plus six differential) and the transition occurs at longer propagation distance for normal spatial patterns while at shorter propagation distance for differential spatial patterns. 
While the traditional MASS WF has a small number of spatial patterns with a discontinuity at $\sim 10\mathrm{km}$, in SH-MASS case, the number of spatial patterns reaches $\sim300$ and their transition distances are continuous from the ground to 20 km high, which implies SH-MASS's aperture geometry gives sufficient number of constraints to estimate a turbulence profile with higher altitude resolution.

\subsection{Turbulence profile reconstruction method}
\label{sec:reconstructionmethod}
Reconstruction of a turbulence profile is solving an inverse problem described as Eq.\ref{eq:nsi=wj} and Eq.\ref{eq:dsi=wj} with analytically-derived WF matrix for observed SIs. 
If we simply apply a linear reconstruction without consideration of the parameter range, negative turbulence strength often appears.
In order to avoid the situation, \citet{tokovinin2003restoration} applies the $\chi^2$ minimization with $y_i$ of $J_i=y_i^2$ as variable. $J_i$ is the turbulence strength of each layer.

In this study, the turbulence profile and associated uncertainty are evaluated based on Bayesian inference with Marcov Chain Monte Carlo (MCMC) method.
As a prior function of MCMC, we applied top-hat filter to limit the parameter space as follows,
\begin{align}
    P(\vec{J}) = 
    \begin{cases}
    1 & \text{if $-32 < \mathrm{log}J_i \mathrm{[m^{1/3}]} < -11$ is satisfied by all $J_i$,} \\
    0 & \text{otherwise,} 
    \end{cases}
\end{align}
where $\vec{J}$ is the turbulence profile. 
The strength range of each turbulence layer i.e. $-32 < \mathrm{log}J_i \mathrm{[m^{1/3}]} < -11$ corresponds to $2.0\times10^{-2} < r_0[\mathrm{m}] < 8.0\times10^{10}$ in the Fried parameter assuming measurement wavelength of $500\ \mathrm{nm}$ and zenith direction. 
It is expected that this parameter range covers the possible turbulence strength of a single layer.
As a likelihood function of MCMC, we use the probability that the observed SIs are obtained from a Gaussian distribution with a mean of the expected SIs and a standard deviation of observation errors as follows,
\begin{align}
    L(\vec{s}|\vec{J}) = \prod_{m=1}^{M} \left( \frac{1}{\sqrt{2\pi \sigma_m^2}} \exp\left[ -\frac{(s_m-(W\vec{J})_m)^2}{2\sigma_m^2} \right] \right),
    \label{eq:LikelihoodFunction}
\end{align}
where $M$ is the number of spatial patterns, $\vec{s}$ and $\vec{\sigma}$ are SIs and their errors, respectively, $W$ is WF matrix, and $\vec{J}$ is the turbulence profile. 
By sampling the multi-dimensional space of the $\vec{J}$ effectively using the Bayesian inference technique, the strength of turbulence and its error is reconstructed for each layer.
The reconstruction procedure was conducted utilizing \textit{emcee}, a MCMC tool for \textit{Python}.

\section{Performance evaluation}
\subsection{Response function}
\label{sec:ResponseFunction}
In order to investigate the SH-MASS's performance of atmospheric turbulence profiling quantitatively, we examine the response of SH-MASS to a single turbulence layer. 
In this calculation, at first, we create a turbulence profile which consists of single turbulence layer at a certain propagation distance. 
Then, we calculate theoretical SIs by multiplying WF to the turbulence profile assuming the measured wavelength of 500 $\mathrm{nm}$. 
The profiling is conducted with a predefined set of layers.
In this study, we select the combinations so that the propagation distance of the lowest layer should be 0.5 km, propagation distance of the highest layer should be 20.0 km and the ratio of the propagation distance of the n-th lowest layer and that of the n+1-th lowest layer should be constant. 
Finally, turbulence profile is reconstructed by using the MCMC method described in Sec.\ref{sec:reconstructionmethod}.

We repeat this procedure with changing the propagation distance of input single turbulence layer in order to obtain a response function, which is defined as estimated turbulence strength of each reconstructed layer as a function of propagation distance of the input single turbulence layer. 
Here, we assume that measurement errors of SIs are 5\% of SIs uniformly for all spatial patterns. 
According to \citet{tokovinin2003restoration}, typical errors of SIs are approximately 2\% except for 3-7\% error for the smallest differential SI.
The size of the error depends on photon flux. 
Thus, our assumption of 5\% errors would be suitable in order to explore the worst case of typical performance of SH-MASS. 

\begin{figure*}
	\includegraphics[width=15cm]{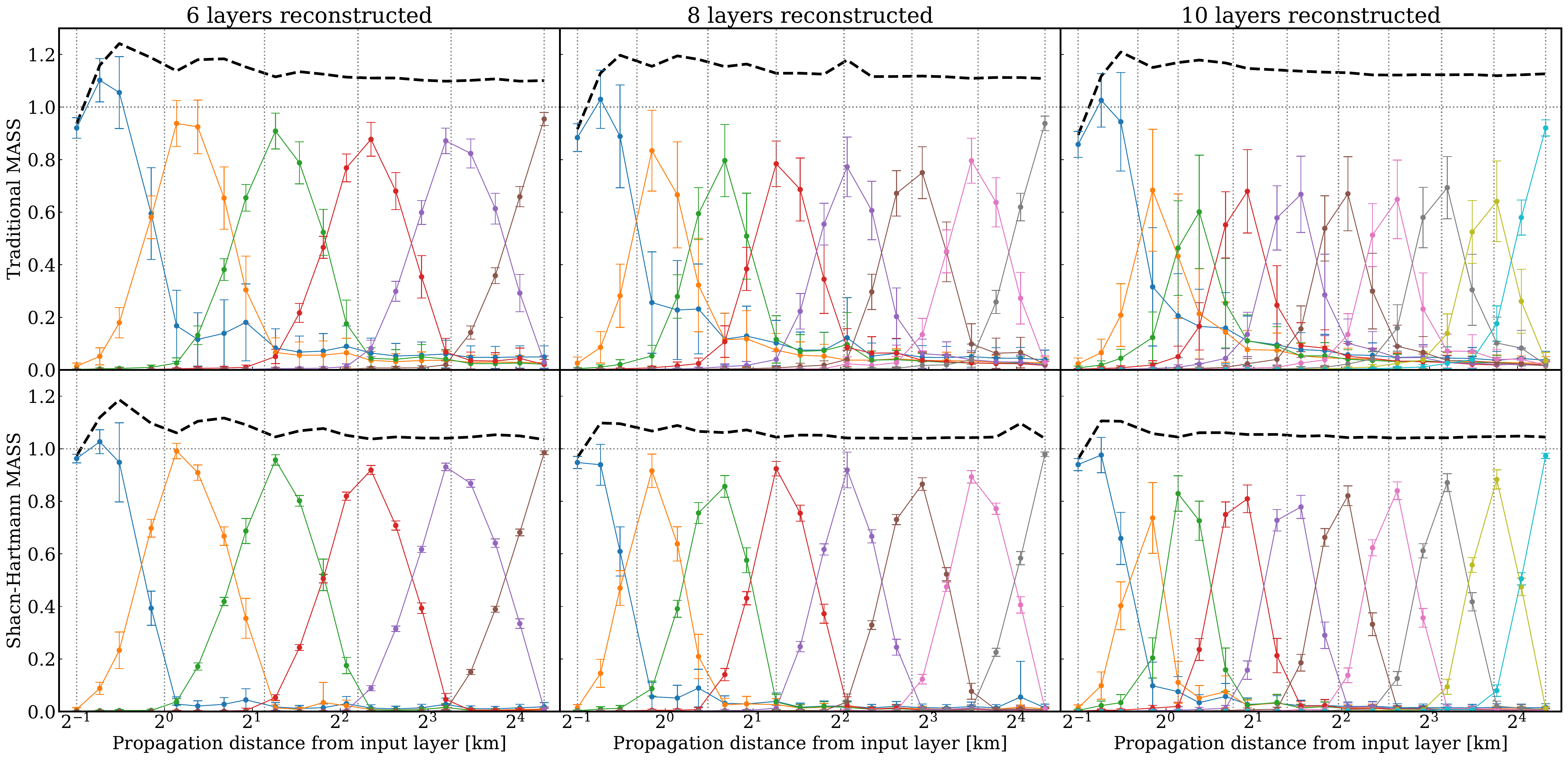}
    \caption{Top panels: The response function for the traditional MASS setup in which turbulence strengths of 6, 8, and 10 layers are reconstructed using scintillation data observed by 4 concentric annular apertures. Bottom panels: The response function for classical SH-MASS setup in which turbulence strengths of 6, 8, and 10 layers are reconstructed using scintillation data observed by 10x10 SH-WFS whose subaperture diameter corresponds to 1.3 $\mathrm{cm}$ on the primary mirror. In each panels, each triangular line 
    represents the response of each reconstruction layer to the input single turbulence layer. In the cases of 6, 8, and 10 layers are reconstructed, the propagation distances from the reconstruction layers are [0.5, 1.0, 2.2, 4.6, 9.6, 20.0] km, [0.5, 0.8, 1.4, 2.4, 4.1, 7.0, 11.8, 20.0] km, and [0.5, 0.8, 1.1, 1.7, 2.6, 3.9, 5.8, 8.8, 13.3, 20.0] km, respectively. Black dashed line represents the total sensitivity as a sum of the response of all the reconstruction layers. Gray vertical lines stand for the propagation distance from reconstructed layers, while gray horizontal line stands for the sensitivity when all input turbulence strength are sensed. }
    \label{fig:ClassicalvsShwfs}
\end{figure*}

\begin{figure}
	\includegraphics[width=\columnwidth]{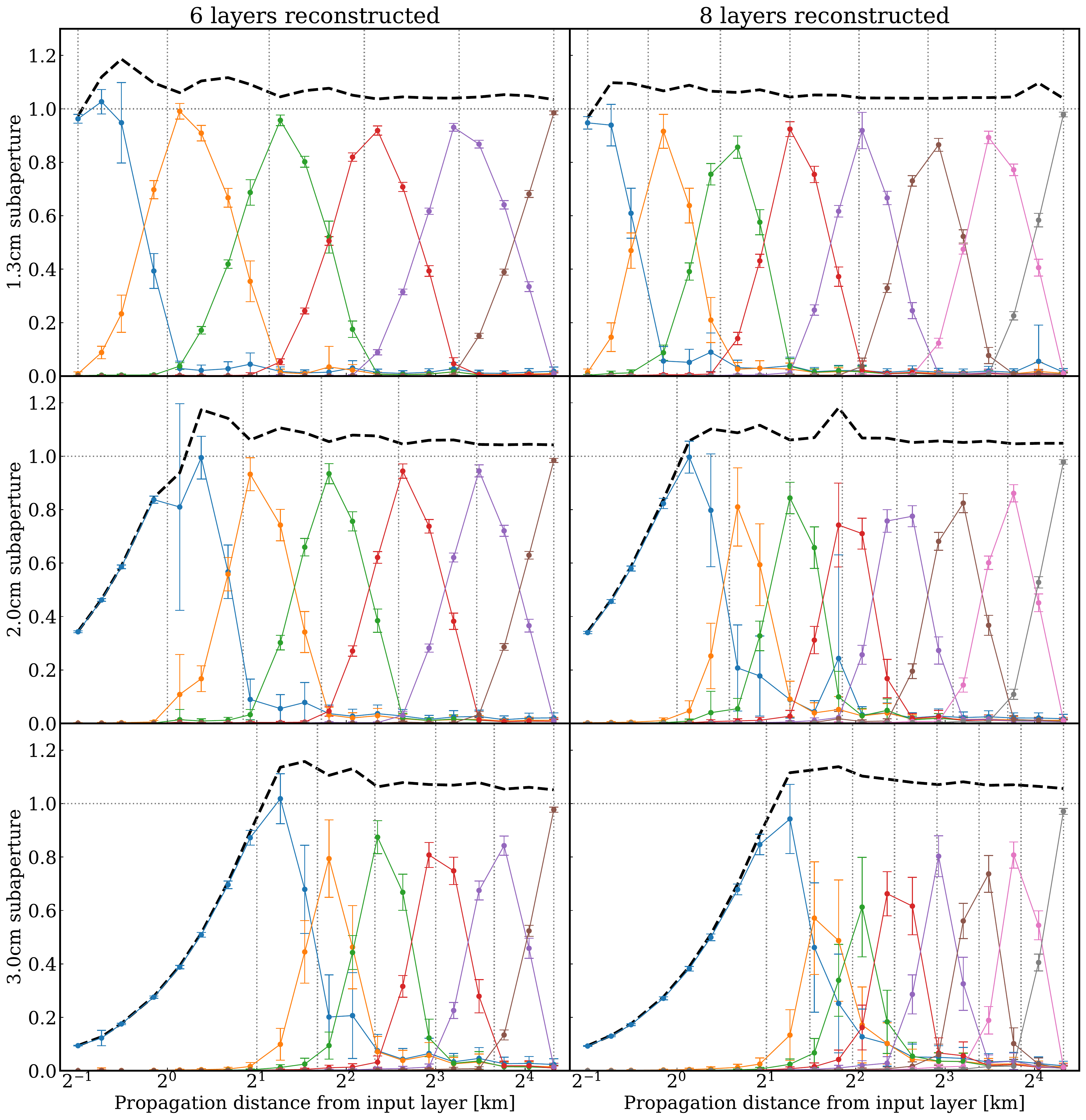}
    \caption{Comparison of the SH-MASS response function for the subaperture diameter of 1.3 $\mathrm{cm}$ (top), 2.0 $\mathrm{cm}$ (middle), and 3.0 $\mathrm{cm}$ (bottom) in the case of 6 (left) and 8 (right) reconstructed layers. Here, the format of SH-WFS is $10\times10$ for all cases. The propagation distance from the lowest reconstruction layer is changed to 1 $\mathrm{km}$ and 2 $\mathrm{km}$ for the subaperture size of 2 $\mathrm{cm}$ and 3 $\mathrm{cm}$, respectively. }
    \label{fig:CompSubapSize}
\end{figure}

\begin{figure}
	\includegraphics[width=\columnwidth]{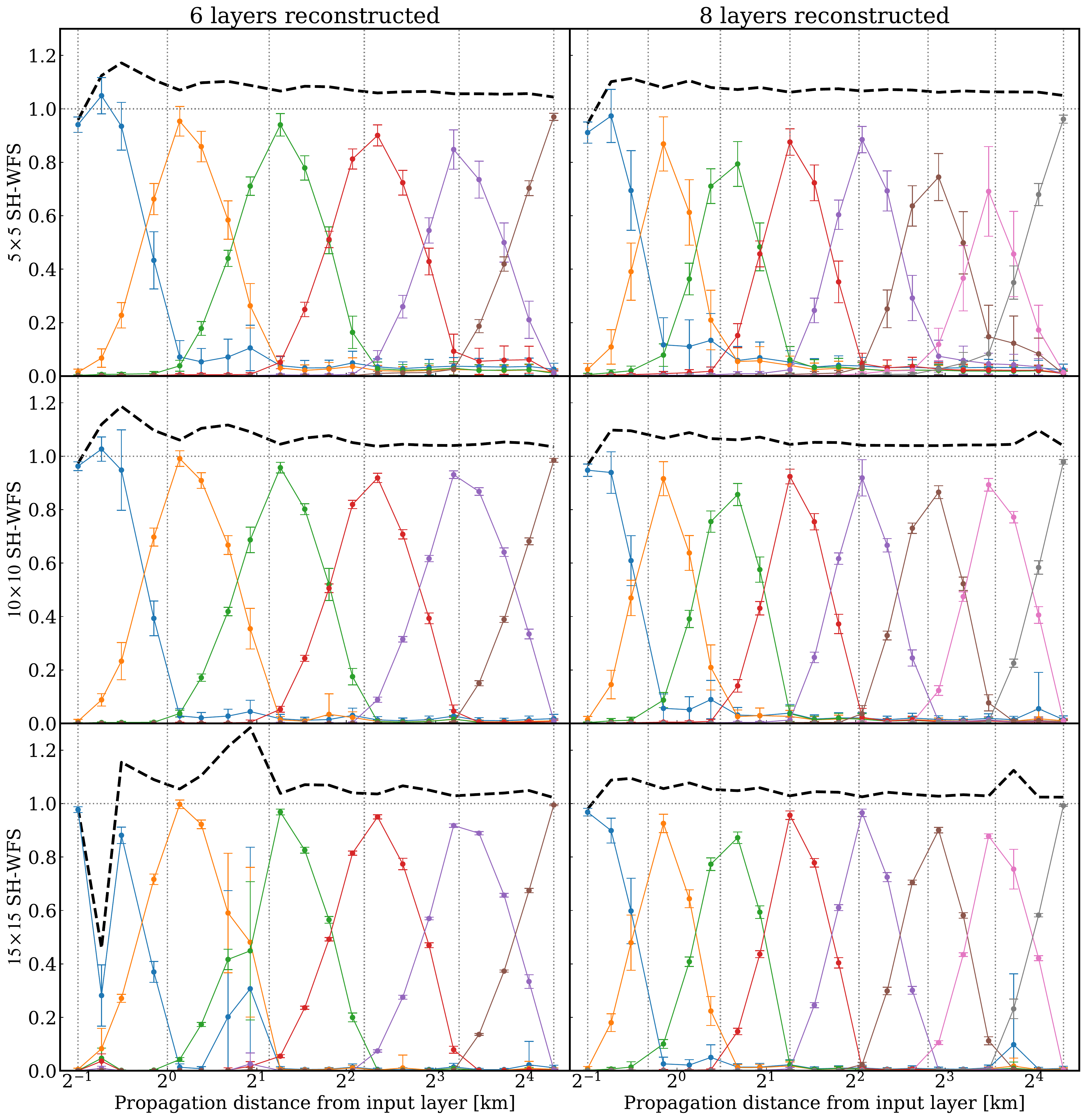}
    \caption{Comparison of the SH-MASS response function for the SH-WFS format of $5\times5$ (top), $10\times10$ (middle), and $15\times15$ (bottom) in the case of 6 (left) and 8 (right) reconstructed layers. Here, the diameter of SH-WFS subaperture is 1.3cm for all cases.}
    \label{fig:CompSubapNumber}
\end{figure}

In Fig.\ref{fig:ClassicalvsShwfs}, the response function of the traditional MASS setup (top panels) and that of SH-MASS setup (bottom panels) are compared. 
The assumed spatial pattern for both of the MASS and SH-MASS are same as Fig.\ref{fig:shmass_aperturepattern} and Fig.\ref{fig:WfComp}.
Different column represents different number of reconstructed layers. In each panel in Fig.\ref{fig:ClassicalvsShwfs}, x-axis represents the propagation distance of input single turbulence layer while y-axis represents the ratio of estimated turbulence strength to the input turbulence strength, hence the sensitivity of MASS to a given propagation distance. 
Each coloured line represents the sensitivity of each reconstructed layer, and errors are defined as the standard deviation of solutions from 10000 MCMC steps after convergence. 
Total sensitivity as a sum of the response of all the reconstruction layers is represented by black dashed line. 

By comparing the top panels with bottom panels in Fig.\ref{fig:ClassicalvsShwfs}, we can see that the size of estimation errors are smaller in SH-MASS. 
In addition, the triangular shape of SH-MASS response function is kept even with the number of reconstructed layers of ten while the triangular shape is broken and the altitude resolution is poorer in the traditional MASS case. 
The black dashed lines are also closer to unity in SH-MASS cases, which means that the estimation of integrated turbulence strength is improved. 
These results indicate that a large number of spatial patterns realized by SH-WFS subaperture geometry is effective in reconstructing turbulence profiles with high altitude resolution and sufficient accuracy.

Fig.\ref{fig:CompSubapSize} and Fig.\ref{fig:CompSubapNumber} show how the shape of response function changes if parameters of SH-WFS are changed. 
In Fig.\ref{fig:CompSubapSize}, the size of subaperture is varied, 1.3 cm, 2.0 cm, and 3.0 cm -diameter cases are shown in top, middle, and bottom panels, respectively. 
Here, format of SH-WFS is fixed to $10\times10$ in all cases. 
If the size of subaperture is increased, fine spatial structure of scintillation is no longer detectable. 
Then, scintillation data does not have any information of turbulence at short propagation distance which is associated with high spatial frequency scintillation pattern. 
For this reason, the propagation distance of lowest reconstructed layer is changed so that the propagation distance $h_{\mathrm{l}}\sec(z)$ and the size of subaperture $x_{\mathrm{subap}}$ satisfy $x_{\mathrm{subap}} \sim \sqrt{\lambda h_{\mathrm{l}}\sec(z)}$, while the propagation distance of the highest reconstructed layer is fixed to $\sim 20\ \mathrm{km}$. 
These figures indicates that the size of subaperture is one of the most important parameter of SH-MASS, which determines the dynamic range of the turbulence profiling. 
Though SH-MASS with small subapertures makes it possible to estimate atmospheric turbulence down to close to the ground, small subaperture can be suffered from the problem of small number of photons. 
The diameter of subaperture should be determined carefully considering the altitude range of estimation, the magnitude of available star, or the availability of highly sensitive detector with high pixel read out rate such as an electron multiplying CCD (EM-CCD).

In Fig.\ref{fig:CompSubapNumber}, the format of SH-WFS is varied, $5\times5$, $10\times10$, and $15\times15$ cases are shown in top, middle, and bottom panels, respectively. 
Here, the diameter of subaperture is fixed to 1.3 cm in all cases. It is clear that the size of error becomes smaller if format of SH-WFS is larger. 
This is because of large number of constraints realized by larger format. 
The number of constraints is 43 (15 normals + 28 differentials), 285 (51 normals + 234 differentials), and 1740 (106 normals + 1634 differentials) for $5\times5$, $10\times10$, and $15\times15$ SH-WFS, respectively. 
On the other hand, in $15\times15$ case, MCMC solver does not converge well in some calculations. 
This would be because too many constraints result in complicated posterior probability distribution and causes poor convergence in Monte Carlo sampling. 
Considering the results and computational cost, the format of $10\times10$ is suitable size of SH-WFS.

\subsection{Required signal to noise ratio}
We investigate the required signal to noise ratio (SNR) in order to conduct the SH-MASS by simulating scintillation observations by Monte Carlo method.
In our calculation, we assume one subaperture of SH-WFS with the diameter of 2.0 cm attached to a general telescope and wavefront sensor system whose total optical transmission is assumed to be $\sim 40\%$. 
We use a band pass of combined V and R bands for photons from a star and background sky assuming a certain magnitude of star, sky brightness of $m_{\mathrm{V,sky}}=21.1\mathrm{[mag\cdot arcsec^{-2}]}$, $m_{\mathrm{R,sky}}=20.6\mathrm{[mag\cdot arcsec^{-2}]}$.
Here, the number of photons are changed stochastically so that it should follow Poisson distribution.
Additionally, in stellar photon case, the effect from scintillation is also considered. 
Assuming SI observed by the subaperture which takes a value from 0.2, 0.5 or 1.0, we randomly vary the number of stellar photon following log-normal distribution (\citealp{zhu2002free}). 
For read-out noise of the detector, we assume stochastic variable which follows a Gaussian distribution with the standard deviation of $1.5\mathrm{[ADU/pixel]}$.
The exposure time is fixed to 2 milliseconds. $n$, the number of photon counting, was taken from  3000, 30000, or 300000 times, which corresponds to 6 seconds, 1 minute, 10 minutes, respectively. 
After $n$-time photon counting, we calculate observed SI and signal to noise ratio (SNR) using the series. 
The definitions of the SI and SNR in this calculation are as follows,
\begin{align}
    \mathrm{SI} &= \frac{\mathrm{Var}[n_{\mathrm{star}}]}{\langle n_{\mathrm{star}}\rangle ^2}, \\
    \mathrm{SNR} &= \frac{\langle n_{\mathrm{star}}\rangle }{\sqrt{\langle n_{\mathrm{star}}\rangle +\mathrm{Var}[n_{\mathrm{sky}}]+\mathrm{Var}[n_{\mathrm{ron}}]}},
\end{align}
where $\langle n_{\mathrm{star}}\rangle $ and $\mathrm{Var}[n_{\mathrm{star}}]$ are average and variance of count from a simulated star, respectively.
$\mathrm{Var}[n_{\mathrm{sky}}]$ is variance of count from background sky, and $\mathrm{Var}[n_{\mathrm{ron}}]$ is variance of count from read-out noise. 
We repeat this Monte Carlo simulation of SI and SNR 100 times, and evaluated the dispersion of the SIs.

\begin{figure}
	\includegraphics[width=\columnwidth]{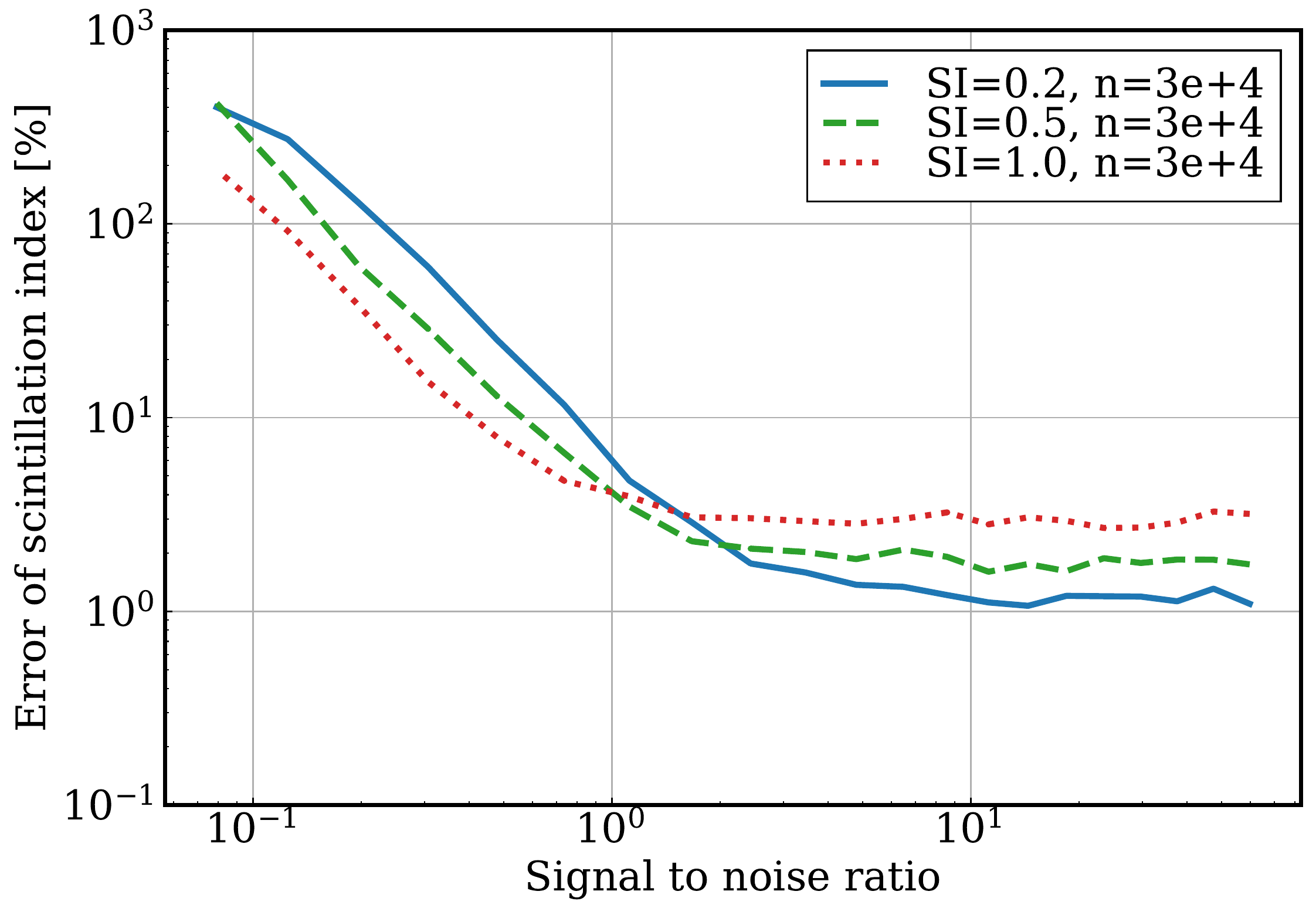}
    \caption{The SI measurement error (the ratio of the standard deviation to the average of 100 measured SIs) as a function of the measured SNR. Lines with different colour means different input SI. As SNR increases, SI measurement error decreases and saturates to a certain value which correlates with real SI value. The saturation occurs at around SNR>3, which would be the required SNR value for accurate SI measurement.}
    \label{fig:snr_sierror1}
\end{figure}

Fig.\ref{fig:snr_sierror1} shows the ratio of standard deviation of the 100 SIs to average of the 100 SIs as a function of SNR. 
In this plot, the number of photon counting is fixed to 30000. Hence one-minute observation for measuring SI is simulated here. 
Lines with different colours correspond to different input SI, and we can see how the SI measurement error decreases when the magnitude of observed star increases depending on the strength of atmospheric turbulence. 
The curves become flat at $\mathrm{SNR}>3$ and the saturated value depends on the input SI. 
Even at strong turbulence condition of $\mathrm{SI}=1.0$, the saturated value is less than 5\%, which means that one-minute observation with $\mathrm{SNR}>3$ is long enough to achieve sufficient SNR assumed in the calculation of the response functions in Sec.\ref{sec:ResponseFunction}. 

\begin{figure}
	\includegraphics[width=\columnwidth]{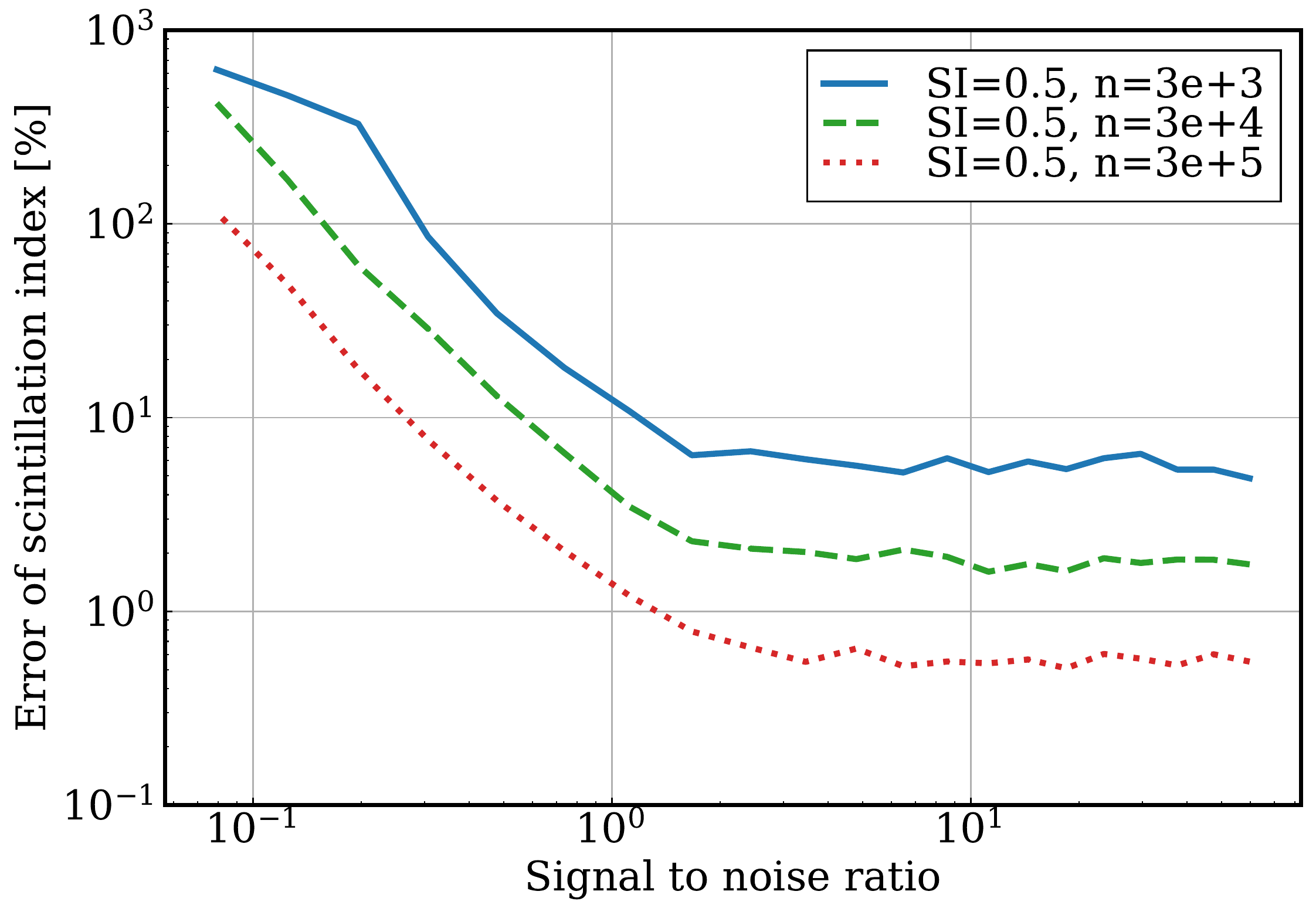}
    \caption{Same with Fig.\ref{fig:snr_sierror1}, but lines with different colour means different number of photon counting. As SNR increases, the SI error decreases and saturates to a certain value which anti-correlates with the number of photon counting. The saturation occurs at SNR>3, which would be the required SNR value for accurate SI measurement.}
    \label{fig:snr_sierror2}
\end{figure}

On the other hand, Fig.\ref{fig:snr_sierror2} is a result with changing $n$ value and fixed input SI of 0.5. 
In this plot also, the ratio of the standard deviation to the average of 100 measured SIs become saturated at SNR higher than $\sim$3. 
The saturated value depends on how many frames are used to estimate the SI. 
This is just because more samples make it possible to more accurately estimate the properties of the SI.
Though more frames make it possible to measure SI with lower measurement error, these frames should be obtained within the characteristic timescale of atmospheric structure evolution. 
Considering the typical timescale is $\sim$ 10 minutes and the result of Fig.\ref{fig:snr_sierror1}, $n=30000$ (1 minute with $500\ \mathrm{Hz}$) will be optimal number for scintillation measurements.

\section{On-sky experiment}
\subsection{Setup and observation}
\begin{figure}
	\includegraphics[width=\columnwidth]{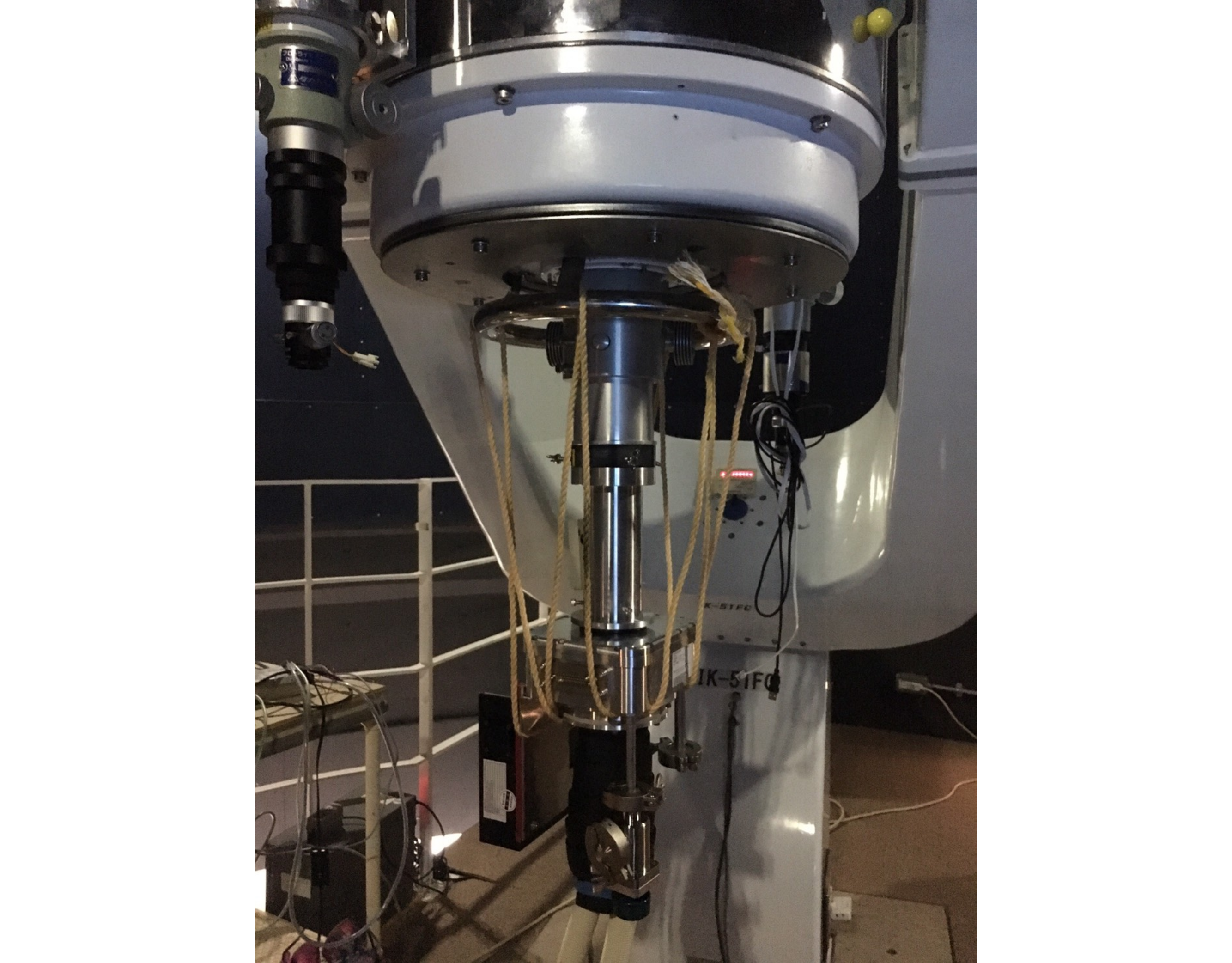}
    \caption{SH-WFS system attached to the 50 cm telescope in Tohoku University. A lenslet array is in the aluminium cylindrical tube at the center of the figure. The dewar attached to the tube is an EMCCD camera. The camera is cooled down using Peltier device and a liquid cooling system.}
    \label{fig:on-sky}
\end{figure}

In order to demonstrate the SH-MASS, we conducted a scintillation observation using a wavefront sensor system attached to the 50 cm telescope, IK-51FC in Tohoku University (see Fig.\ref{fig:on-sky}).
Our wavefront sensor system is consist of a collimator, a Bessel's R band filter, whose central wavelength is 630 $\mathrm{nm}$, a 150 $\mathrm{\mu m}$ pitch lenslet array (Thorlabs, MLA150-5C) which has a chromium mask for blocking light that reaches outside of the circular aperture of each microlens, relay lenses and an EMCCD camera with E2V CCD60 $128\times128$ 24 $\mathrm{\mu m}$ pixel detector and custom made readout electronics provided by Nuvu Cameras. 
The primary mirror of 50 cm telescope is effectively divided into $20\times20$ by the lenslet array, in other words, the effective diameter of a subaperture correspond to 2.5 $\mathrm{cm}$ on the primary mirror. 
The field of view of each subaperture is $\sim 35$ arcsec.
The detector was cooled down to $-30^\circ \mathrm{C}$ so that dark current noise can be negligible. 
One pixel of the detector corresponds to 4.9 arcsec on the sky. 
Amplification signal of 42.6 V is applied to achieve factor 300 multiplication gain of the EMCCD.
High speed imaging of $500\mathrm{Hz}$ was repeated 30000 times targeting Deneb ($m_R=1.14[\mathrm{mag}]$). 
This procedure was repeated nine times in one hour on a clear night, October 16th, 2019 in Japan Standard Time.
In the one hour, the elevation of the star changed from \ang{46} to \ang{34}.

\subsection{Data analysis}
First step is measurement of count value of each SH-WFS spot in each frame.
The spot reference frame is created by averaging the 30000 frames in each dataset, and spot size and locations of spots are measured. 
By fitting each spot of the reference frame with a Gaussian function, the diameter of Airy disk is measured to be 4.45 pixels (FWHM is 1.88 pixels).
Then, for each frame, we define inside of circles with a center of each spot location and a diameter of 5.0 pixels as spot region and others as background region.
Background count is estimated as mean count of the background region.
After background subtraction from all the pixels, each spot count is calculated as the total counts of each circle in the spot region.
By this procedure, background-subtracted spot counts are calculated for all the spots in the 30000 frames.

Count fluctuation seen in the 30000 frames is ascribed to photon noise and atmospheric scintillation.
Variance of the fluctuation contributed from photon noise is the mean photon count, while that contributed from scintillation is proportional to the square of the mean photon count.
Therefore, considering the observed photon count of $\sim 100$, contribution from scintillation is the dominant component.
According to \citet{zhu2002free}, the distribution of light intensity induced by scintillation follows log-normal distribution. 
Then, we check if the histograms of spot counts follow a log-normal distribution written as follows.
\begin{align}
    f(x) = \frac{A}{\sqrt{2\pi}\sigma x}\mathrm{exp}\left( \frac{(\mathrm{ln}x-\mu)^2}{2\sigma^2} \right),
\end{align}
where $\mu$ and $\sigma$ are shape parameters of the distribution and $A$ is a normalization parameter.
\begin{figure}
	\includegraphics[width=\columnwidth]{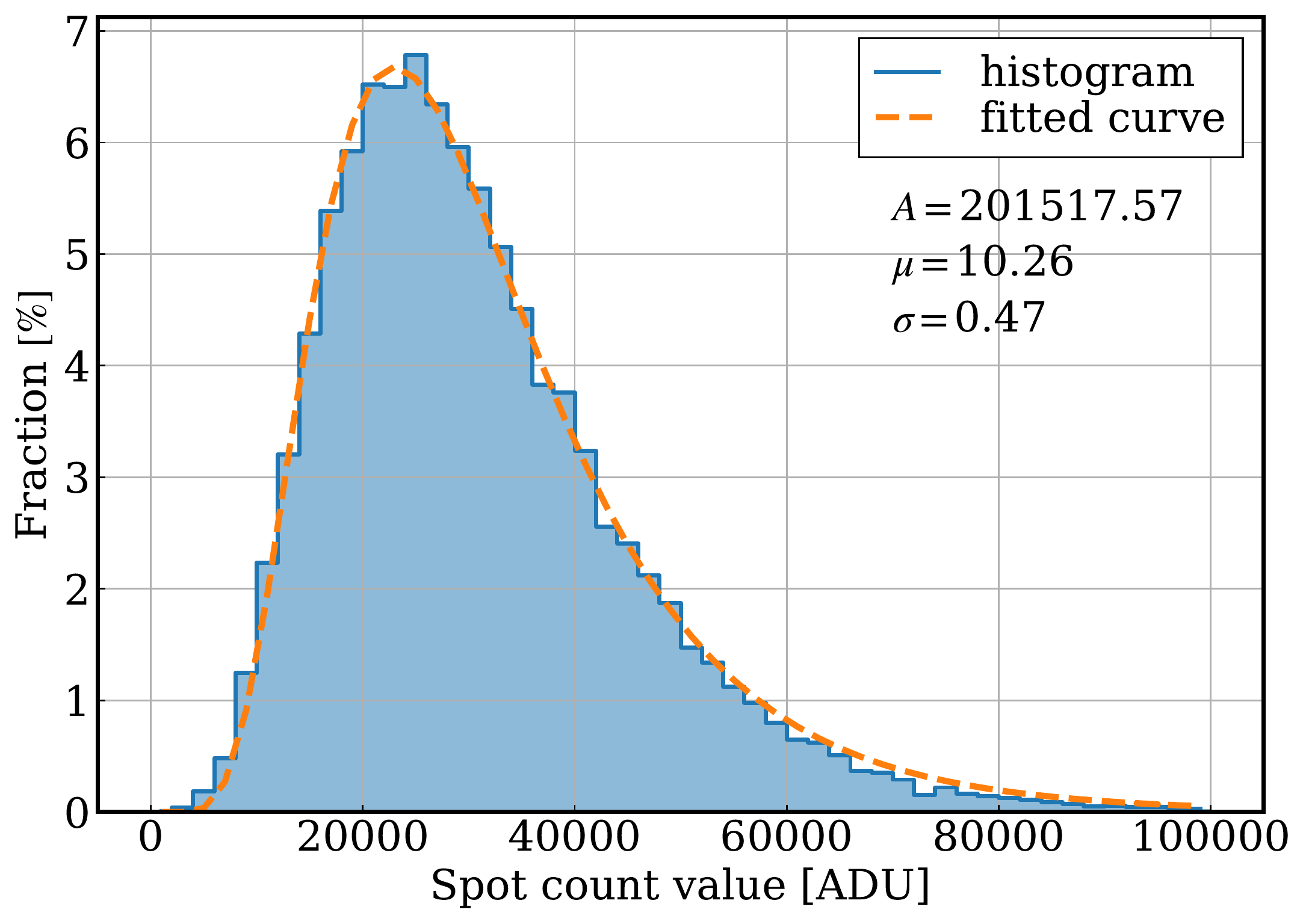}
    \caption{Histogram of count values of one spot of SH-WFS measured with 500 Hz in 1 minute. Horizontal axis represents photon counts after multiplication by the EM-CCD.}  The distribution is well fitted by a log-normal distribution, which implies that the detected intensity fluctuation is scintillation.
    \label{fig:lognormal}
\end{figure}
Fig.\ref{fig:lognormal} shows the histogram of count values of a spot in 1 minute. 
The histogram is well fitted by a log-normal distribution function with parameters of $A=2.0\times10^5$, $\mu=10.26$ and $\sigma=0.47$. 
All other spots also follow similar shape of the histogram, which supports that the observed intensity fluctuation of SH-WFS spots are caused by atmospheric turbulence. 
However, it should be noted that the log-normal distribution of the count values is not a sufficient condition for concluding that the fluctuation is due to the scintillation.

Then, SIs of all the spatial patterns are calculated from the count fluctuations of spots. The mean, variance and covariance of each spot's count fluctuation are computed and SIs are calculated following Eq.\ref{eq:nsi_spotstat} and Eq.\ref{eq:dsi_spotstat}.

Finally, effects from finite exposure time are corrected. 
In this observation, the images are obtained with 2 milliseconds exposure, which means that fluctuation components which have the timescale of less than 4 milliseconds is smoothed out.
Therefore, real SIs which can ideally observed by 0 millisecond exposure time should be larger than that measured by the above-mentioned procedure. 
In this study, we follow the method described in \citet{tokovinin2003restoration} in which 0 millisecond SIs are estimated from linear extrapolation of SIs measured by $\tau$ milliseconds exposure and that measured by $2\tau$ milliseconds exposure,
\begin{align}
    s_{0} = 2s_{\tau} - s_{2\tau}.
\end{align}
Here, datasets with $2\tau$ milliseconds exposure is effectively obtained by averaging two adjacent images in the data of $\tau$ milliseconds exposure.

\begin{figure}
	\includegraphics[width=\columnwidth]{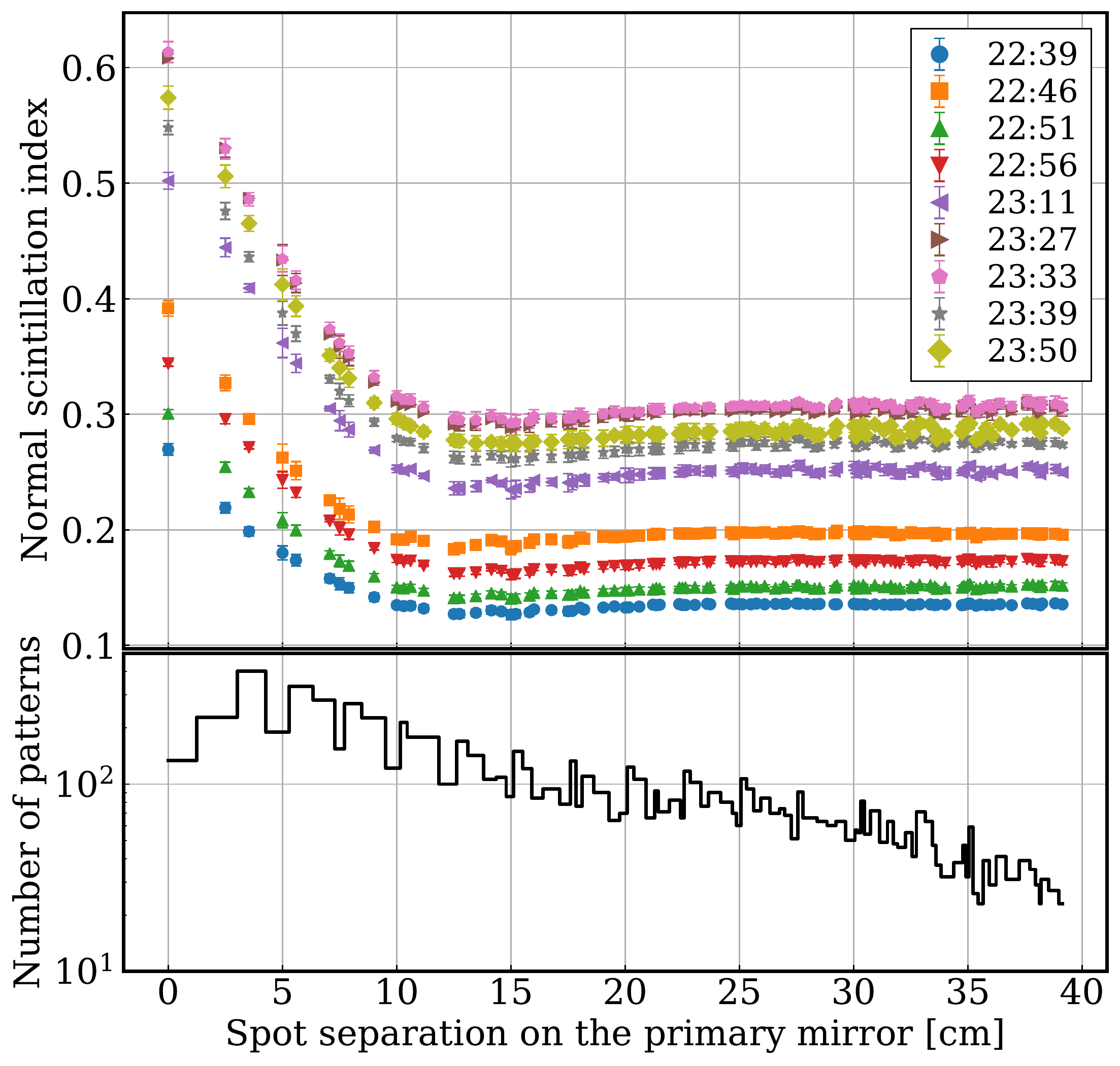}
    \caption{Top: The observed normal SIs are plotted as a function of a separation distance of two subapertures which constitutes a spatial pattern. Different colour represents different observation time. The value of SIs decreases and become flattened as the separation increases, which indicates that the typical correlation length of scintillation is shorter than $15\ \mathrm{cm}$. And the trend that SIs become larger as time goes by can be explained by variation of the elevation angle of the star Bottom: The number of subaperture pairs which have common separation distance. The normal SIs and their errors in top figure have been calculated as the mean and standard deviation of these number of statistics.}
    \label{fig:nsi}
\end{figure}

Top panel of Fig.\ref{fig:nsi} shows the observed normal SIs as a function of separation of two subapertures which constitutes a normal spatial pattern. 
Each colour of plots represents the difference of observed time. 
The error on each normal SI is small enough to discern the difference between each scintillation state of each observation time. 
Bottom panel of Fig.\ref{fig:nsi} shows the number of subaperture pairs which have common spatial pattern. 
At all observation time, the normal SI decreases as a function of subaperture separation and get flattened at spatial length of 10-15 cm and longer. 
This feature reflects that there are little atmospheric turbulence at higher than $\sim 20\ \mathrm{km}$. 
In fact, the spatial scale of 10-15 cm is consistent with the typical spatial scale of scintillation created by turbulence layer i.e. $\sqrt{\lambda h \sec(z)}\sim 13.4\  \mathrm{cm}$ with the assumption of wavelength $\lambda \sim 600\  \mathrm{nm}$, altitude $h \sim 20\ \mathrm{km}$, and airmass $\sec(z) \sim 1.5$.
In addition, this feature can be understood using Eq.\ref{eq:nsi_spotstat}. 
For null separation case, $\mathrm{Cov}[I_i,I_j]$ becomes $\mathrm{Var}[I_i]$, and SI becomes $\mathrm{Var}[I_i]/\langle I_i\rangle ^2$. Whereas for very long separation case, $\mathrm{Cov}[I_i,I_j]$ becomes 0, and SI becomes $\mathrm{Var}[I_i]/2\langle I_i\rangle ^2$, half of the SI for the null separation case. In Fig.\ref{fig:nsi} actually, normal SIs for longer separation than 15 cm is almost half of normal SI for 0 cm separation.
Besides, there is a trend that the value of SI become larger as time goes by. 
This can be explained by the change of the elevation of the star. 
As the elevation becomes lower, apparent altitude of turbulence layer becomes higher. 
Both effects account for the increase of SIs. 
These properties also support that detected fluctuation of stellar intensity is due to the atmospheric turbulence.

\subsection{Atmospheric turbulence profile}
\begin{figure}
	\includegraphics[width=\columnwidth]{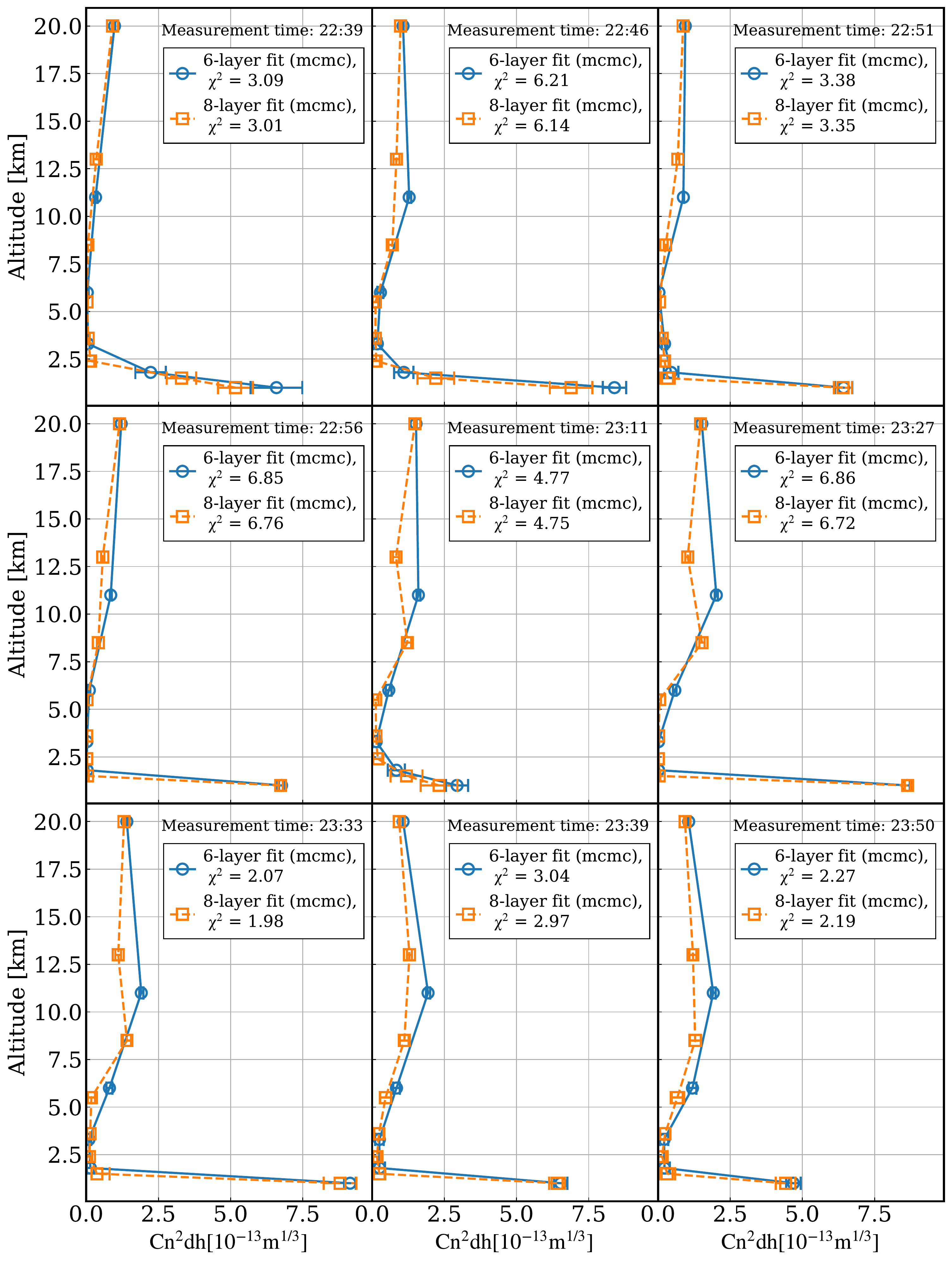}
    \caption{The atmospheric turbulence profile reonstructed by SIs measured at Tohoku university. These profiles are reconstructed by the MCMC estimation method mentioned in Sec.\ref{sec:reconstructionmethod}. The effect of elevation angle of the star is corrected. Different panel corresponds to different observation time which is described in each panel's title. Blue lines are a profiles which are reconstructed assuming six layers while orange lines assume eight layers. Reduced $\chi^2$ values of each profile estimation are shown in the legend. }
    \label{fig:tp1}
\end{figure}

Fig.\ref{fig:tp1} shows the atmospheric turbulence profile reconstructed from the observed SIs by the MCMC estimation method described in Sec.\ref{sec:reconstructionmethod}. 
Different panel corresponds to different observation time which is described in each panel's title.
The propagation distances of reconstructed layers are varied so that the reconstruction altitudes of the layers should be same among observation times.
Blue lines are profiles which are estimated assuming six layers (altitude of 1.0, 1.8, 3.3, 6.0, 11.0, 20.0 km above the telescope aperture) while orange lines assume eight layers (altitude of 1.0, 1.5, 2.4, 3.6, 5.5, 8.5, 13.0, 20.0 km above the telescope aperture). 
The reduced $\chi^2$ values which are less than 10 in all datasets and the small uncertainties which represent one sigma values of turbulence strengths after MCMC convergence implies that the observed normal and differential SIs are described well with the scintillation model. 
In addition, the overall shape of the profiles shows that the strongest turbulence exists at the lowest layers and second strongest peak distributes at roughly 10 km.
The profiles are consistent with that expected from the typical characteristics of the Earth's atmosphere such as the ground turbulent layer and the tropopause, respectively. 

\section{Discussions}
\subsection{Fast profile reconstruction using iterative method}
\label{sec:FastProfile}
Although MCMC-based profile estimation method enables us to evaluate the estimation error, it requires large calculation cost, which typically takes a few tens of minutes for six-layer reconstruction with eight-core parallel processing using \textit{Intel}\textregistered\  \textit{Core}\texttrademark\ \textit{i7-4790K} CPU and highly depends on the number of reconstructed layers.
On the other hand, atmospheric turbulence profile as a prior information for tomographic reconstruction matrix has to be updated in a timescale of tens of minutes. 
Then, we try faster profile calculation based on Broyden–Fletcher–Goldfarb–Shanno (BFGS) algorithm, which is one of iterative solvers for non-linear optimization problem. 
This algorithm can be utilized with \textit{scipy.optimize.minimize} module for \textit{Python}. 
We impose the same condition of $-32<\mathrm{log}J_i\mathrm{[m^{1/3}]}<-11$ for all the component of $\vec{J}$ as MCMC-based method, and minimize the $\chi^2$ function directly.

However, in iterative calculation method, the solution is not necessarily the global minimum.
Actually, in the current case, one-time iterative calculation does not give an identical solution.
Hence, we conduct the BFGS algorithm 1000 times from 1000 different random initial turbulence profile. 
Then, we pick out 100 final turbulence profiles whose $\chi^2$ values are the smallest and calculated the mean and standard deviation of the 100 profiles.

\begin{figure}
	\includegraphics[width=\columnwidth]{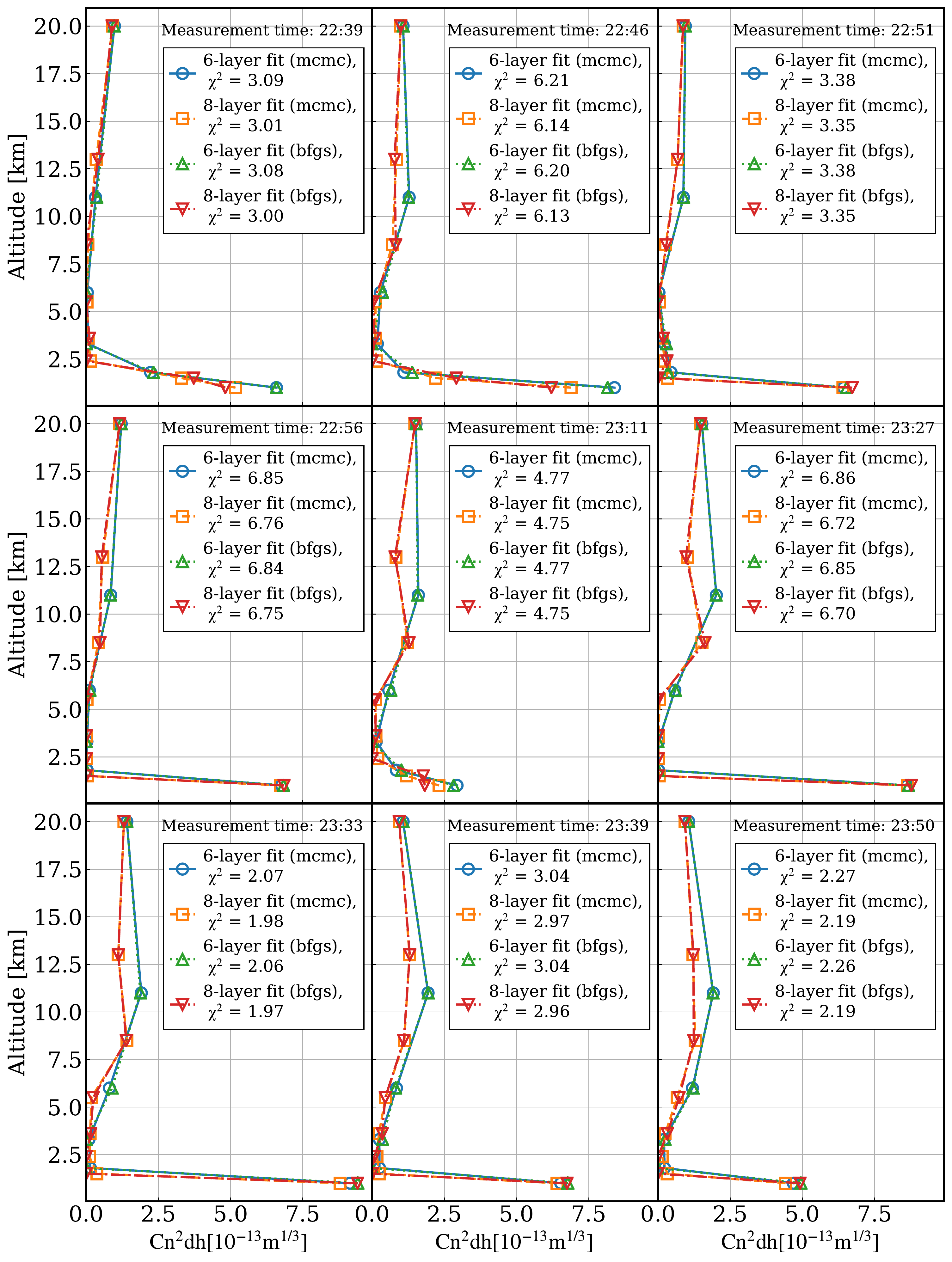}
    \caption{Comparison of atmospheric turbulence profile reconstructed by the MCMC based method (blue and orange lines) and iterative method (green and red lines). The effect of elevation angle of the star is corrected. It can be understood that iterative method reproduces same profile as the MCMC method. Different panel corresponds to different observation time which is described in each panel's title. Reduced $\chi^2$ values of solutions from each method are shown in the legend.}
    \label{fig:tp2}
\end{figure}

\begin{figure}
	\includegraphics[width=\columnwidth]{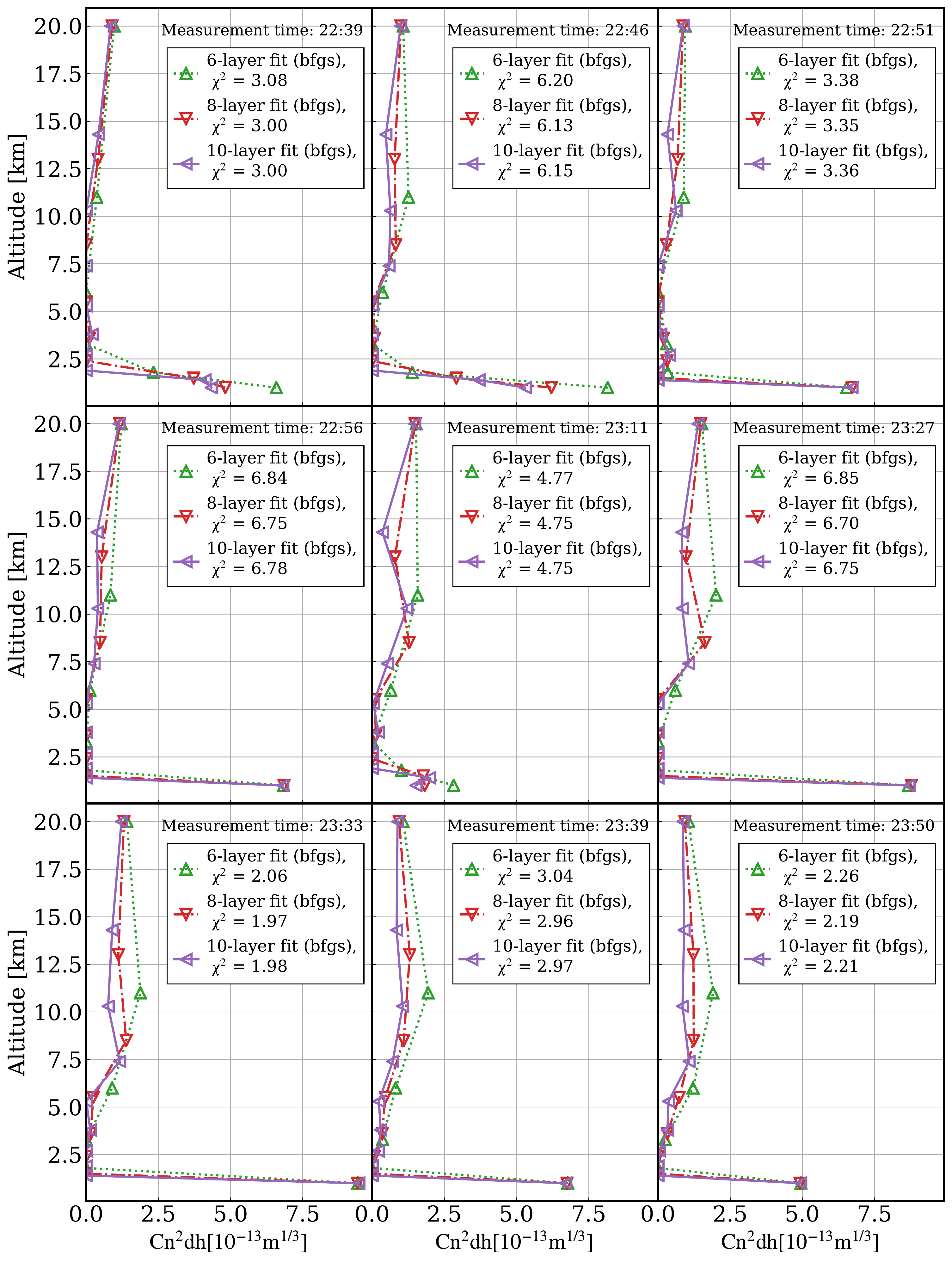}
    \caption{Atmospheric turbulence profile at Tohoku university. These profiles are reconstructed from the observed SIs by the iterative estimation method mentioned in Sec.\ref{sec:FastProfile}. The effect of elevation angle of the star is corrected. Different panel corresponds to different observation time which is described in each panel's title. Green, red and purple lines are a profiles which are reconstructed assuming six, eight and ten reconstructed layers, respectively. Reduced chi squared values of each profile estimation are described in the legend.}
    \label{fig:tp3}
\end{figure}

In Fig.\ref{fig:tp2}, we compare the turbulence profile obtained by BFGS iterative method with that obtained by the MCMC method. 
In all observation time, both estimation methods reproduce the same turbulence profile. 
The consistency suggests that 1000-time iterative minimization from the random initial profiles is sufficient to find out the global minimum.
Because the calculation time for the 1000-time iterative minimization is typically a few minutes, the iterative BFGS method can be used for a faster profile reconstruction. 
Then, we conduct the profile estimation with 10 layers (1.0, 1.4, 1.9, 2.7, 3.8, 5.3, 7.4, 10.3, 14.3, 20.0 km), which takes the timescale of days when MCMC-based method is used.
The result is shown in Fig.\ref{fig:tp3}. 
Here, higher altitude resolution with $dh/h=1.4$ ($dh/h=2.0$ for the traditional MASS) is realized for atmospheric turbulence which distributes from 1.0 km to 20.0 km. 
By increasing the number of reconstruction layers, it turns out to be that the strong turbulence seen at 11.0 km in the 6-layer fitting result consists of turbulent layers which distribute extended altitudes around 7.5-15 km.
Precise understanding of turbulence distribution realized by the high altitude resolution is necessary to produce a realistic reconstruction matrix.

\subsection{Low sensitivity to lower altitude}
As mentioned in previous works (e.g. \citealp{avila1997whole}), turbulence at less than several hundred meter high is undetectable by scintillation-based profiling methods. 
This is because the variance of observed intensity, or scintillation is proportional to the propagation distance with the power of 5/6th.
This characteristic can be seen in our result of response function (Fig.\ref{fig:ClassicalvsShwfs}, Fig.\ref{fig:CompSubapSize} and Fig.\ref{fig:CompSubapNumber}) in which the estimation error of turbulence strength becomes larger for shorter propagation distance of input layer.
In order to overcome this limitation, three solutions can be considered as future improvement plans.
The first is generalized mode. 
Likewise G-SCIDAR, turbulence at low altitude can be measured by placing SH-WFS at some distance away from the pupil plane. 
The second is DIMM mode. In DIMM, two image motions of single bright star is measured by two apertures whose center separation is typically 20-30 cm. 
By using SH-WFS spots, star image motion is observed by various separation length. 
Therefore, DIMM can be conducted by using SH-WFS system and turbulence strength at ground layer can be estimated by the difference between DIMM measurement and MASS measurement.
The third solution is a combination with SLODAR. 
SLODAR uses two SH-WFSs to measure the correlation of wavefront distortion from a double star. 
Because triangulation-based profiler like SLODAR does not have any sensitivity to turbulence at high altitude, some SLODAR system is optimized for profiling of the ground layer \citep{butterley2020characterization}. 
Combining SH-MASS with SLODAR enables us to profile whole atmospheric turbulence using a single optical system.

\section{Summary}
In this study, we investigate a new MASS-based atmospheric turbulence profiling method called SH-MASS, which reproduces the profile from scintillation observed by a Shack-Hartmann wavefront sensor. 
By evaluating the response function of the SH-MASS in comparison with those of the traditional MASS, it is shown that SH-MASS theoretically has higher altitude resolution than traditional MASS under the assumption that the scintillation measurements have 5\% error.
This high altitude resolution is enabled by a large number of spatial patterns realized by the grid pattern of SH-WFS. 

By investigating the behaviour of response functions with changing the parameters of SH-MASS, the larger size of subaperture makes lower sensitivity to the low altitude turbulence, therefore smaller size of subaperture is better as long as the signal to noise ratio of each spot in a SH-WFS image is larger than $\sim 3$. 

This new profiler is demonstrated with 50 cm telescope at Tohoku University and typical characteristics of the atmospheric turbulence are reproduced as the estimated turbulence profile. 
In order to decrease the calculation cost and meet the real time requirement of the profiling, i.e., one profile estimation per $\sim 10$ minutes, we confirm that faster iterative method can also reproduce the same profile as the MCMC-based method. 

\section*{Acknowledgements}
The authors thank Drs. Yosuke Minowa, Kazuma Mitsuda, and Koki Terao for helpful discussions. HO is supported by Graduate Program on Physics for the Universe (GP-PU), Tohoku University. MA is supported by JSPS KAKENHI (17H06129). Part of this work was achieved using the grant of Joint Development Research supported by the Research Coordination Committee, National Astronomical Observatory of Japan (NAOJ), National Institutes of Natural Sciences (NINS).

\section*{Data availability}
The data underlying this article will be shared on reasonable request to the corresponding author.



\bibliographystyle{mnras}
\bibliography{bibliography} 

\begin{thebibliography}{}
\makeatletter
\relax
\def\mn@urlcharsother{\let\do\@makeother \do\$\do\&\do\#\do\^\do\_\do\%\do\~}
\def\mn@doi{\begingroup\mn@urlcharsother \@ifnextchar [ {\mn@doi@}
  {\mn@doi@[]}}
\def\mn@doi@[#1]#2{\def\@tempa{#1}\ifx\@tempa\@empty \href
  {http://dx.doi.org/#2} {doi:#2}\else \href {http://dx.doi.org/#2} {#1}\fi
  \endgroup}
\def\mn@eprint#1#2{\mn@eprint@#1:#2::\@nil}
\def\mn@eprint@arXiv#1{\href {http://arxiv.org/abs/#1} {{\tt arXiv:#1}}}
\def\mn@eprint@dblp#1{\href {http://dblp.uni-trier.de/rec/bibtex/#1.xml}
  {dblp:#1}}
\def\mn@eprint@#1:#2:#3:#4\@nil{\def\@tempa {#1}\def\@tempb {#2}\def\@tempc
  {#3}\ifx \@tempc \@empty \let \@tempc \@tempb \let \@tempb \@tempa \fi \ifx
  \@tempb \@empty \def\@tempb {arXiv}\fi \@ifundefined
  {mn@eprint@\@tempb}{\@tempb:\@tempc}{\expandafter \expandafter \csname
  mn@eprint@\@tempb\endcsname \expandafter{\@tempc}}}

\bibitem[\protect\citeauthoryear{Arsenault et~al.,}{Arsenault
  et~al.}{2012}]{arsenault2012eso}
Arsenault R.,  et~al., 2012, in Adaptive Optics Systems III. p. 84470J

\bibitem[\protect\citeauthoryear{Avila, Vernin  \& Masciadri}{Avila
  et~al.}{1997}]{avila1997whole}
Avila R.,  Vernin J.,   Masciadri E.,  1997, Applied Optics, 36, 7898

\bibitem[\protect\citeauthoryear{Beckers}{Beckers}{1988}]{beckers1988increasing}
Beckers J.~M.,  1988, in European Southern Observatory Conference and Workshop
  Proceedings. p.~693

\bibitem[\protect\citeauthoryear{Butterley, Wilson, Sarazin, Dubbeldam, Osborn
  \& Clark}{Butterley et~al.}{2020}]{butterley2020characterization}
Butterley T.,  Wilson R.,  Sarazin M.,  Dubbeldam C.,  Osborn J.,   Clark P.,
  2020, Monthly Notices of the Royal Astronomical Society, 492, 934

\bibitem[\protect\citeauthoryear{Costille \& Fusco}{Costille \&
  Fusco}{2012}]{costille2012impact}
Costille A.,  Fusco T.,  2012, in Adaptive Optics Systems III. p. 844757

\bibitem[\protect\citeauthoryear{Farley, Osborn, Morris, Fusco, Neichel,
  Correia  \& Wilson}{Farley et~al.}{2020}]{farley2020limitations}
Farley O.,  Osborn J.,  Morris T.,  Fusco T.,  Neichel B.,  Correia C.,
  Wilson R.,  2020, Monthly Notices of the Royal Astronomical Society, 494,
  2773

\bibitem[\protect\citeauthoryear{Fusco \& Costille}{Fusco \&
  Costille}{2010}]{fusco2010impact}
Fusco T.,  Costille A.,  2010, in Adaptive Optics Systems II. p. 77360J

\bibitem[\protect\citeauthoryear{Gendron, Morel, Osborn, Martin, Gratadour,
  Vidal, Le~Louarn  \& Rousset}{Gendron et~al.}{2014}]{gendron2014robustness}
Gendron E.,  Morel C.,  Osborn J.,  Martin O.,  Gratadour D.,  Vidal F.,
  Le~Louarn M.,   Rousset G.,  2014, in Adaptive Optics Systems IV. p. 91484N

\bibitem[\protect\citeauthoryear{Gilles, Wang  \& Ellerbroek}{Gilles
  et~al.}{2008}]{gilles2008wavefront}
Gilles L.,  Wang L.,   Ellerbroek B.,  2008, in Adaptive Optics Systems. p.
  701520

\bibitem[\protect\citeauthoryear{Guesalaga, Perera, Osborn, Sarazin, Neichel
  \& Wilson}{Guesalaga et~al.}{2016}]{guesalaga2016fass}
Guesalaga A.,  Perera S.,  Osborn J.,  Sarazin M.,  Neichel B.,   Wilson R.,
  2016, in Adaptive Optics Systems V. p. 99090H

\bibitem[\protect\citeauthoryear{Guesalaga, Ayanc{¥'a}n, Sarazin, Wilson,
  Perera  \& Le~Louarn}{Guesalaga et~al.}{2020}]{guesalaga2020fass}
Guesalaga A.,  Ayanc{¥'a}n B.,  Sarazin M.,  Wilson R.,  Perera S.,
  Le~Louarn M.,  2020, Monthly Notices of the Royal Astronomical Society

\bibitem[\protect\citeauthoryear{Hammer et~al.,}{Hammer
  et~al.}{2004}]{hammer2004falcon}
Hammer F.,  et~al., 2004, in Second backaskog workshop on extremely large
  telescopes. pp 727--736

\bibitem[\protect\citeauthoryear{Kornilov, Tokovinin, Vozyakova, Zaitsev,
  Shatsky, Potanin  \& Sarazin}{Kornilov et~al.}{2003}]{kornilov2003mass}
Kornilov V.,  Tokovinin A.~A.,  Vozyakova O.,  Zaitsev A.,  Shatsky N.,
  Potanin S.~F.,   Sarazin M.~S.,  2003, in Adaptive Optical System
  Technologies II. pp 837--845

\bibitem[\protect\citeauthoryear{Kornilov, Tokovinin, Shatsky, Voziakova,
  Potanin  \& Safonov}{Kornilov et~al.}{2007}]{kornilov2007combined}
Kornilov V.,  Tokovinin A.,  Shatsky N.,  Voziakova O.,  Potanin S.,   Safonov
  B.,  2007, Monthly Notices of the Royal Astronomical Society, 382, 1268

\bibitem[\protect\citeauthoryear{Lardi{¥`e}re et~al.,}{Lardi{¥`e}re
  et~al.}{2014}]{lardiere2014multi}
Lardi{¥`e}re O.,  et~al., 2014, in Adaptive Optics Systems IV. p. 91481G

\bibitem[\protect\citeauthoryear{Marchetti et~al.,}{Marchetti
  et~al.}{2007}]{marchetti2007sky}
Marchetti E.,  et~al., 2007, The Messenger, 129

\bibitem[\protect\citeauthoryear{Minowa et~al.,}{Minowa
  et~al.}{2017}]{minowa2017ultimate}
Minowa Y.,  et~al., 2017, Adaptive Optics for Extremely Large Telescopes V
  (AO4ELT5)],(Jun 2017)

\bibitem[\protect\citeauthoryear{Rigaut}{Rigaut}{2002}]{rigaut2002ground}
Rigaut F.,  2002, in European Southern Observatory Conference and Workshop
  Proceedings. p.~11

\bibitem[\protect\citeauthoryear{Rigaut \& Neichel}{Rigaut \&
  Neichel}{2018}]{rigaut2018multiconjugate}
Rigaut F.,  Neichel B.,  2018, Annual Review of Astronomy and Astrophysics, 56,
  277

\bibitem[\protect\citeauthoryear{Rigaut et~al.,}{Rigaut
  et~al.}{2014}]{rigaut2014gemini}
Rigaut F.,  et~al., 2014, Monthly Notices of the Royal Astronomical Society,
  437, 2361

\bibitem[\protect\citeauthoryear{Rocca, Roddier  \& Vernin}{Rocca
  et~al.}{1974}]{rocca1974detection}
Rocca A.,  Roddier F.,   Vernin J.,  1974, JOSA, 64, 1000

\bibitem[\protect\citeauthoryear{Sarazin \& Roddier}{Sarazin \&
  Roddier}{1990}]{sarazin1990eso}
Sarazin M.,  Roddier F.,  1990, Astronomy and Astrophysics, 227, 294

\bibitem[\protect\citeauthoryear{Saxenhuber, Auzinger, Le~Louarn  \&
  Helin}{Saxenhuber et~al.}{2017}]{saxenhuber2017comparison}
Saxenhuber D.,  Auzinger G.,  Le~Louarn M.,   Helin T.,  2017, Applied Optics,
  56, 2621

\bibitem[\protect\citeauthoryear{Stone, Hu, Mills  \& Ma}{Stone
  et~al.}{1994}]{stone1994anisoplanatic}
Stone J.,  Hu P.,  Mills S.,   Ma S.,  1994, JOSA A, 11, 347

\bibitem[\protect\citeauthoryear{Tallon \& Foy}{Tallon \&
  Foy}{1990}]{tallon1990adaptive}
Tallon M.,  Foy R.,  1990, Astronomy and Astrophysics, 235, 549

\bibitem[\protect\citeauthoryear{Tokovinin}{Tokovinin}{2003}]{tokovinin2003polychromatic}
Tokovinin A.~A.,  2003, JOSA A, 20, 686

\bibitem[\protect\citeauthoryear{Tokovinin}{Tokovinin}{2004}]{tokovinin2004seeing}
Tokovinin A.,  2004, Publications of the Astronomical Society of the Pacific,
  116, 941

\bibitem[\protect\citeauthoryear{Tokovinin, Kornilov, Shatsky  \&
  Voziakova}{Tokovinin et~al.}{2003}]{tokovinin2003restoration}
Tokovinin A.,  Kornilov V.,  Shatsky N.,   Voziakova O.,  2003, Monthly Notices
  of the Royal Astronomical Society, 343, 891

\bibitem[\protect\citeauthoryear{Vidal, Gendron  \& Rousset}{Vidal
  et~al.}{2010}]{vidal2010tomography}
Vidal F.,  Gendron E.,   Rousset G.,  2010, JOSA A, 27, A253

\bibitem[\protect\citeauthoryear{Wilson}{Wilson}{2002}]{wilson2002slodar}
Wilson R.~W.,  2002, Monthly Notices of the Royal Astronomical Society, 337,
  103

\bibitem[\protect\citeauthoryear{Zhu \& Kahn}{Zhu \& Kahn}{2002}]{zhu2002free}
Zhu X.,  Kahn J.~M.,  2002, IEEE Transactions on communications, 50, 1293

\makeatother
\end{thebibliography}








\bsp	
\label{lastpage}
\end{document}